\documentclass[12pt]{iopart}
\pdfoutput=1
\usepackage{iopams}
\usepackage{graphicx, color}

\begin{document}

\title{Gaussian Lattice Boltzmann method and its applications to rarefied flows}

\author{Oleg  Ilyin}

\address{Dorodnicyn Computing Centre of Russian Academy of Sciences, Vavilova st. 40, 119333 Moscow, Russia}
\ead{oilyin@gmail.com}
\vspace{10pt}

\begin{abstract}
 A novel discretization approach for the  Bhatnager-Gross-Krook (BGK) kinetic equation is  proposed. A hierarchy of LB models starting from $D1Q3$ model with increasing  number of velocities converging  to BGK model is derived. The method inherits properties of the Lattice Boltzmann (LB) method like linear streaming step, conservation of moments. Similar to the finite-difference methods for the BGK model the presented approach describes high-order moments of the distribution function with controllable error.  The Sod shock tube problem, the Poiseuille flow between parallel plates and the plane Couette flow are considered for wide range of Knudsen numbers. Good stability and significant increase in precision  over the conventional LB models are observed.
\end{abstract}

\vspace{2pc}

\section{Introduction}
Nowadays the Lattice Boltzmann (LB) approach 
\cite{2015Succi_epl, 2016Succi_aip, 2018succi} is supposed to be an useful tool in modeling  of  non-equilibrium rarefied flows  \cite{2005toschi, 2006zhang, 2007ansumali,2007niu,2008kim,2008tang, 2008Yudistiawan, 2009verhaege, 2010Yudistiawan,2011meng, 2014ambrus,2015mont,2016feucht_schleif, 2017silva}. 
Nevertheless, when the flow is rarefied and the role of high-order moments of the velocity distribution function increases the precision of conventional LB models go down.

In the conventional LB method the discretization is  performed  in a such  way that the LB distribution function  is equivalent  to the solution for the BGK equation projected on a finite basis in a velocity space spanned by the Hermite polynomials (the Grad expansion). This equivalence  is achieved via the  Gauss-Hermite quadratures \cite{1997he, 1998shan, 2006shan, 2010shan}. Since  the first moments of the LB local equilibrium are reproduced in the same form as for the local Maxwell state then LB method is conservative  by the construction.  
The application of an additional regularization procedure \cite{Latt2006, Chen2006, Mattila2017} guarantees that the solutions of LB  will be confined in this finite velocity space and  results in increased stability and  precision of the LB models \cite{2006zhang, 2007niu, 2014mont, 2014mont2, 2015mont}.

Instead of the exact reproduction of the  first moments for the  local Maxwell state, the  finite-difference  methods for the BGK model are aimed to recover the distribution function in overall with decreasing error when the discrete velocity set is growing \cite{2000mieussens}. Potentially all the moments of  the distribution function are recovered but with some error.
The cost to pay is the usage of relatively large discrete velocity sets, moreover   conservation of the first moments requires some additional efforts and the streaming step does not have concise form like in the LB method.

The present research is aimed to develop a new LB based discrete velocity (DV) method which is able to cope with the kinetic high-order moments in the rarefied flow but without drastic increase in number of discrete velocities.  The starting point  of the  presented construction is a well-known one-dimensional three-velocity $D1Q3$ LB model \cite{2017kruger}.  At the  next step, a summation procedure for the LB model is  introduced: an addition of $D1Q3$ model to $D1Q3$ model  yields $D1Q5$ model. The  repetition of the summation $k$ times  leads to  models with $(3+2k)$ velocities. All the models in the  hierarchy have the same order of the isotropy   equals  to the  order of  the isotropy  of the root model ($D1Q3$), moreover the models are calibrated at the same flow velocity and temperature. The  models in the  hierarchy  have monotonically increasing precision in a following sense. The  magnitudes of the  errors in all highest moments (in the comparison with the   local Maxwell state) are  uniformly  decreasing  when $k$ is growing. Using the Central Limit Theorem (CLT)  one can show  that the hierarchy of LB models converges to the BGK model.  The  presented method covers advantageous  properties of the both DV and LB  methods.
The  Gaussian shape of the local equilibrium state is better reproduced in  each subsequent summation step which results in better reproduction of the half-moments and the  kinetic boundary conditions. 

The numerical experiments presented in Section 5  support the theoretical predictions.
The models show good stability in the Sod shock tube  problem. For the  the  plane Couette flow and Poiseuille flow between parallel walls the convergence to the benchmark solutions is  observed.

\section{The Construction of the Hierarchy of Lattice Boltzmann models and Central Limit Theorem}

The simplest LB  one-dimensional discretization  of the BGK equation  which is able to reproduce the Navier-Stokes equations at the limit of low Mach numbers $Ma$ (with $O(Ma^3)$ error)
is D1Q3 model \cite{2017kruger}
$$
f_{i}(t+\Delta t, x+c_i\Delta t)-f_{i}(t,x)=\frac{\Delta t}{\tau+\frac{\Delta t}{2}}\left(f^{eq}_{i}-f_{i} \right)(t,x),
$$
where $\tau$ is the relaxation time,  $f_{i}, i=-1,0,1$ are the lattice distribution functions
corresponding  to the lattice velocities $c_i: -c, 0, +c$, here $c\Delta t$ is the distance  between the lattice  nodes, $\Delta t$ is the  lattice  time step. 
The form of a  collision frequency $\frac{1}{\tau+\frac{\Delta t}{2}}$
guarantees second-order accuracy in physical space \cite{2017kruger}.
The  equilibrium states $f^{eq}_{\pm 1},f^{eq}_{0}$  are defined as
$$
f^{eq}_{\pm 1}(t,x)=\frac{\rho(t,x)}{6}\left(1\pm3\frac{u(t,x)}{c}+3\frac{u(t,x)^2}{c^2}\right),
$$
$$
f^{eq}_{0}(t,x)=\frac{4\rho(t,x)}{6}\left(1-3\frac{u(t,x)^2}{2c^2}\right),
$$
where the macroscopic  parameters $\rho, u$ are defined from the expressions
$$
\rho(t,x)=f_{-1}(t,x)+f_0(t,x)+f_{+1}(t,x),
$$
$$
\rho(t,x)u(t,x)=-f_{-1}(t,x)c+f_{+1}(t,x)c,
$$
moreover the full energy is calculated as follows
$$
\rho(t,x)(u(t,x)^2+c_s^2)=(f_{-1}(t,x)+f_{+1}(t,x))c^2,
$$
where the sound velocity is fixed:  $c_s=\sqrt{\frac{1}{3}}c$, i.e. the model describes only isothermal flows. 
This  model serves as a root  of the presented below hierarchy.

Now lets associate with the  local equilibrium distribution function  for $D1Q3$ model a random variable $X(N)$,  here $N=1$ (the first step in the  hierarchy). Assume  that this random variable has  three outcomes  $-c(1),0,+c(1)$, where $c(1)$ is some lattice velocity. Assume that $X(1)$  has  the distribution function same as  the  local equilibrium distribution for $D1Q3$ model. 
At the  present  moment  this  procedure is formal and  does not give any additional information. The  underlying reason for the  introduction of the random variable $X(1)$ will be clear at the  next step.

Next, consider two independent identically distributed (i.i.d.) random variables  $X_1(2), X_2(2)$  (here $N=2$ , the second step) where  each of the variables  has again three outcomes $-c(2),0,+c(2)$ in a such way that their sum  has  the same expected value (the  bulk velocity $u$) and  the same variance (the temperature $c_s^2$)
as for $N=1$ case.

Now, generalizing  the  previous  step  consider  a sum  of $N$ i.i.d. random variables (each has three outcomes $-c(N),0,+c(N)$) such that their sum  has  an
expected value $u$ and variance $c_s^2$.
The variable $\sum_j^N X_j(N)$  has $2N+1$ possible outcomes
$-Nc, (-N+1)c \ldots Nc,  \, c\equiv c(N)$.
Since  this  sum is composed of the independent identical random variables  then the corresponding distribution function can  be calculated  in an exact closed form. 
As a result, a triangular array of the random variables can be constructed
$$
X(1),
$$
$$
X_1(2), X_2(2),
$$
$$
X_1(3), X_2(3), X_3(3),
$$
$$
\ldots \ldots \ldots \ldots
$$
$$
X_1(2), X_2(2), \ldots \ldots X_N(N).
$$
The Central Limit  Theorem (CLT) for triangular arrays in the form of  Lyapunov or  Lindeberg-Feller \cite{1996ferguson} guarantees  that the sequence of the distribution functions in the  hierarchy converges to the Gaussian distribution (the Maxwell state with the bulk velocity $u$ and the temperature $c_s^2$).  

It will be convenient to define as $ Prob\left(\sum_{j=1}^N X_j=nc\right)$ a  probability for a sum of $N$ random variables $X_j(N), j=1 \ldots N$ to have  a value $nc(N), n=-N \ldots N$.  

The evaluation of $Prob\left(\sum_{j=1}^N X_j=nc\right)$ can be reduced  to the following  problem. Assume that a  dice is rolled $N$ times, each dice roll has three outcomes $0, \pm 1$ with the  probabilities $ p_{0},p_{\pm}$. One  needs  to find the probability that a sum of rolls take the value $n, |n| \leq N $.  
Assume  that  $ n \geq 0$ (the case of negative $n$ can be  considered in a similar way) and the value $-1$  is obtained  $m$ times. Then the value $1$ is  rolled $n+m$ times, and  the  value $0$ is  obtained $N-(n+m)-m \geq 0$ times. The  latter inequality  means that $m$ is  lesser than $(N-m)/2$.
Since  $m$ can take  only integer values then  $m$ lies  in the  interval from $0$ to $ \lfloor\frac{N-n}{2}\rfloor$. 
The  number  of ways to get $-1$  result $m$ times and  $1,0$ results $n+m, N-n-2m$ times respectively is $N!/(n+m)!m!(N-n-2m)!$. Then the required  probability  is $\sum_{m=0}^{\lfloor\frac{N-n}{2}\rfloor} \frac{N!}{(n+m)! m! (N-n-2m)!}p_{+}^{n+m} p_{-}^{m} p_{0}^{N-n-2m}$ .

Finally, at $N$-s step ($N-1$ summations) the following $2N+1$ discrete velocity model is introduced
for $-Nc, (-N+1)c \ldots Nc, \, c\equiv c(N)$ lattice 
\begin{equation}\label{maineq01}
f_n(t+\Delta t, x+nc \Delta t)- f_n(t,x)=
\frac{\Delta t}{\tau+\frac{\Delta t}{2}}\left(f^{eq}_{n;N}-f_n  \right)(t,x), 
\end{equation}
where $n= -N\ldots N$ and 
\begin{equation}\label{maineq02}
f^{eq}_{n;N}(t,x)=\rho Prob\left(\sum_{j=1}^NX_j=nc\right),
\end{equation}
$$
Prob\left(\sum_{j=1}^NX_j=nc\right)=
$$
\begin{equation}\label{maineq03}
=\sum_{m=0}^{\lfloor\frac{N-n}{2}\rfloor} \frac{N!}{(n+m)! m! (N-n-2m)!}P_{N,+}^{n+m} P_{N,-}^{m} P_{N,0}^{N-n-2m}
\end{equation}
for $ n\geq0$ ,  where $\lfloor\cdot \rfloor$ is the rounding to lowest integer and
\begin{equation}\label{maineq04}
P_{N,\pm}=\frac{1}{2c^2N}\left(c_s^2 \pm cu+\frac{u^2}{N}\right), 
\end{equation}
\begin{equation}
P_{N,0}=1-P_{N,+}-P_{N,-}
\end{equation}
and
\begin{equation}\label{maineq05}
c_s^2= \frac{Nc^2}{3},
\end{equation}
where  for the sake of brevity the shortened notation for the lattice  velocity $c$ is  used instead of $c(N)$.
To keep the temperature constant $c_s^2=\theta_0$ at the  all levels of the  hierarchy it should be required 
\begin{equation}\label{maineq06}
c \equiv c(N)=\sqrt{3\theta_0/N},
\end{equation}
thus the lattice step is  decreasing when $N$ grows. 
Similar expressions  can be obtained for
$n<0$ by taking $|n|$  instead of $n$ 
and changing $P_{N,\pm}$ by $P_{N, \mp}$.
For the sake of clarity an example for $N=2$ case is  addressed in Appendix A. One can convince that the  shape of the equilibrium states $f^{eq}_{n;N}$ readily converges to Gaussian after a few summation steps.

The  model (\ref{maineq01})-(\ref{maineq05}) is  the  main result of the  paper. It can considered as a new LB type discretization for the BGK kinetic  equation (for isothermal flows).  

\section{Analytical properties}
The equilibrium states  in the hierarchy contain the functions $P_{N,\pm}, P_{N,0}$. This states are non-negative if $P_{N,\pm}, P_{N,0}$ are non-negative. This requirement leads to  the following  inequality $ |u| \leq  \sqrt{\frac{2}{3}}Nc $ or 
$ |u| \leq  \sqrt{2N\theta_0} $.
Therefore, the domain of non-negativity is  growing when $N$ increases. Potentially  this property can result  in better stability for the  models with $N>1$ in comparison with the conventional $D1Q3$ model. 

The  most  interesting question is the  reproduction of the highest moments of the local Maxwell state by the presented method.
By a straightforward computation one can convince that the difference  between the third   moment for the local Maxwell distribution and  the third  moment for $N$-s  model in the  hierarchy is
$$
\frac{u^3}{N^2}.
$$
 The  order of isotropy is constant in the  hierarchy but the overall magnitude of the error is decreasing as $N^{-2}$.
This is very similar  to the finite-difference methods  for BGK model, for which the errors appear in all moments but they are suppressed at some rate when the number  of discrete velocities is growing. 

The  fourth moment behaves in a  similar  way. For the difference  between the local Maxwell fourth moment and the fourth moment for the local equilibrium states in the  hierarchy one has the expression
$$
\frac{\theta_0u^2}{N^2}-\left(\frac{4}{N^2}+\frac{3}{N^3}  \right)u^4,
$$
and again similarly to the $D1Q3$ model all the models  in the hierarchy have  $O(u^2)$ leading error term. The amplitude of the errors decrease as $N^{-2}$.

The straightforward computation of the moment generating function (MGF) defined by $M(s)$ is complicated. One has
$$
M(s) \equiv <e^{ns}> = \sum_{n=-N}^N e^{ns} Prob\left(\sum_{j=1}^NX_j=nc\right).
$$
The convolution of the sums seems  to be problematic. Nevertheless, the result can be  obtained  if  one  takes  in account the fact that the  local equilibrium is related  to a sum $Y=\sum^N_j X_j(N)$ of  independent random variables $X_j(N)$. Then the  moment generating function for $Y$ is a  product of the moment generating functions for $X_j(N)$, they are defined below as $M_X(s)$. 
As a result
\begin{equation}\label{mgf}
M(s)=M_X(s)^N=(P_{N,-}e^{-cs}+P_{N,0}+P_{N,+}e^{cs})^N,
\end{equation}
then any moment $m_k$ of  order $k$ can be calculated from Eq. (\ref{mgf}) using the formula 
$$ 
m_k=\frac{d^k M(s)}{ds^k}|_{s=0}.
$$
Now taking logarithm from the MGF function (\ref{mgf}) and using the expressions (\ref{maineq04})  one  obtains  the following expression 
$$
\log(M(s))=\frac{\theta s^2}{2}+us+
$$
$$
+\left \{(9/6!-1/16)s^6+\ldots
+(-u^3/2)s^3  \right \}\frac{1}{N^2}+O\left(\frac{1}{N^3} \right),
$$
where  $\frac{\theta s^2}{2}+us $ is  the  logarithm of MGF
for the Gaussian distribution and $\theta$ is some constant temperature. Therefore, the  difference between the logarithms  of MGF for the  LB local equilibrium state in Eqs. (\ref{maineq01})-(\ref{maineq05})  and  the local Maxwell state is  of 
$ O(1/N^2)$ order. Then one concludes that  the  moments  for
the presented  LB models converge to the local Maxwell ones  with the error decreasing as $O(1/N^2)$. 

\section{Models  in several dimensions and  ballistic streamers removal}

The models  in several dimensions can  be constructed as a tensor product of 1D models. It will be convenient to denote the models based on the  presented summation procedure as $G$-$DaQb$ (Gaussian LB model in $a$ dimensions with $b$ velocities).
Since all the  models  in the discussed above  hierarchy  have the same  order  of isotropy ($D1Q3$, $G$-$D1Q5$, $G$-$D1Q7$ and so on) then
one can construct  2D and  3D  models in the form
$G$-$D1Qn\times G$-$D1Qm$ and $G$-$D1Qn\times G$-$D1Qm \times G$-$D1Qk$ with $n*m$, $n*m*k$  velocities respectively, the order of isotropy for the multidimensional models will be the same as in 1D case. The  local equilibrium takes the product form of the equilibrium states for 1D models.  This  product form approach is based  on the ideas from \cite{2010Karlin}.  

For instance, the simplest multidimensional model (2D case)  is $15$ velocity  model $G$-$D2Q15$ composed by
the $G$-$D1Q5$ (the formal sum of $D1Q3$ and $D1Q3$) and $D1Q3$ model.

This models have  velocities  parallel to the axis - a streaming directions which do not collide with a wall if  the wall is placed parallel to the axis (ballistic streamers effect) \cite{2005toschi}. The removal of zero lattice velocity  mitigates the problem and significantly increases the calculation precision for several problems \cite{2005toschi, 2016feucht_schleif}.

Having $(2N+1)$ velocity model  for $[-Nc \ldots 0 \ldots Nc]$ lattice the transformation to  $(2N+2)$ velocity model (zero velocity is removed) for $[-(N+0.5)c,  -(N-0.5)c, \ldots (N-0.5)c, (N+0.5)c]$ lattice is proposed. Here the lattice step $c$ is  not fixed by the relation (\ref{maineq06}) since another relation between the temperature and  lattice velocity will be  obtained.  This model  reads as
$$
f_{n+\frac{1}{2}}\left(t+\Delta t, x+\left(n+\frac{1}{2}\right)c \Delta t \right)- f_{n+\frac{1}{2}}(t,x)=
$$
\begin{equation*}
=\frac{\Delta t}{\tau+\frac{\Delta t}{2}} \left(f^{eq}_{n+\frac{1}{2};N}-f_{n+\frac{1}{2}}  \right)(t,x), \quad  n= 0\ldots N, 
\end{equation*}
$$
f_{n-\frac{1}{2}}\left(t+\Delta t, x+\left(n-\frac{1}{2}\right)c \Delta t \right)- f_{n-\frac{1}{2}}(t,x)=
$$
\begin{equation*}
=\frac{\Delta t}{\tau+\frac{\Delta t}{2}} \left(f^{eq}_{n-\frac{1}{2};N}-f_{n-\frac{1}{2}}  \right)(t,x), \quad  n= -N, 
\end{equation*}
where $f_{n+\frac{1}{2}},f_{n+\frac{1}{2}; N}^{eq}$  are the distribution functions and local equilibrium states for the lattice velocities $(n+\frac{1}{2})c, \, n=0 \ldots N$; $f_{n-\frac{1}{2}},f_{n-\frac{1}{2}; N}^{eq}$  are the distribution functions and local equilibrium states for the lattice velocities $(n-\frac{1}{2})c, \, n=-N \ldots 0$.

Here the values  of $f_{\pm n \pm \frac{1}{2}; N}^{eq}$  are taken as the average value of the local distribution states from $2N+1$ hierarchy for the neighbouring to $(\pm n \pm \frac{1}{2})c$ lattice velocities $\pm nc$ and $\pm (n+1)c$
\begin{equation}\label{maineq01_noball}
f_{n+\frac{1}{2};N}^{eq}\left(nc+\frac{1}{2}c\right)=\frac{1}{2}\left\{f_{n+1;N}^{eq}(nc+c)+f_{n;N}^{eq}(nc)\right\}, 
\end{equation}
valid for $0\leq n<N$  and
\begin{equation}\label{maineq02_noball}
f_{N+\frac{1}{2};N}^{eq}\left(Nc+\frac{1}{2}c\right)=\frac{1}{2}f_{N;N}^{eq}(Nc),
\end{equation}
also
\begin{equation}\label{maineq03_noball}
f_{n-\frac{1}{2};N}^{eq}\left(nc-\frac{1}{2}c\right)=\frac{1}{2}\left\{f_{n-1;N}^{eq}(nc-c)+f_{n;N}^{eq}(nc)\right\}
\end{equation}
valid for $-N< n \leq 0$ and
\begin{equation}\label{maineq04_noball}
f_{-N-\frac{1}{2};N}^{eq}\left(-Nc-\frac{1}{2}c\right)=\frac{1}{2}f_{-N;N}^{eq}(-Nc),
\end{equation}
where $f_{n,N}^{eq}$ are the local equilibrium states from $2N+1$ velocities  hierarchy (\ref{maineq04}).
One can convince  that 
$$
\sum_{n=-N}^0 f_{n-\frac{1}{2};N}^{eq}+\sum_{n=0}^N f_{n+\frac{1}{2};N}^{eq}=
\sum_{n=-N}^{N} f_{n;N}^{eq}=\rho
$$
and
$$
\sum_{n=-N}^0 f_{n-\frac{1}{2};N}^{eq}\left[nc-\frac{1}{2}c\right]+\sum_{n=0}^N f_{n+\frac{1}{2};N}^{eq} \left[nc+\frac{1}{2}c\right] =
$$
$$
=\sum_{n=-N}^{N}f^{eq}_{n;N} nc=\rho u.
$$
The second  moment  for the $2N+2$ hierarchy is  shifted
$$
\sum_{n=-N}^0 f_{n-\frac{1}{2};N}^{eq}\left[nc-\frac{1}{2}c\right]^2+\sum_{n=0}^N f_{n+\frac{1}{2};N}^{eq} \left[nc+\frac{1}{2}c\right]^2= 
$$
$$
=\rho \frac{c^2}{4}+\sum_{n=-N}^{N}f_{n;N}^{eq} (nc)^2=\rho \frac{c^2}{4}+ \rho u^2+\rho \frac{Nc^2}{3} ,
$$
therefore the models $2N+2$ velocities are calibrated at the  temperature $\theta$ which is given by the following expression
\begin{equation}\label{temp_2N2}
\theta=\frac{Nc^2}{3}+\frac{c^2}{4}=\frac{4N+3}{12}c^2,
\end{equation}
the constant temperature $\theta=\theta_0$ is required at the all levels of the hierarchy then  the following formula for the  lattice velocities should be applied   
$$
c \equiv c(N)=\sqrt{\frac{12}{4N+3}\theta_0}.
$$
Finally, it is worth to mention that the error terms in the
third  moment for $2N+2$ hierarchy (\ref{maineq01_noball})-(\ref{maineq04_noball}) in comparison to the Maxwell  distribution are the same as for $2N+1$ hierarchy (\ref{maineq01})-(\ref{maineq05})
(Appendix B.).

\begin{figure*}
 \begin{minipage}[t]{.99\textwidth}
        \includegraphics[width=0.45\textwidth]{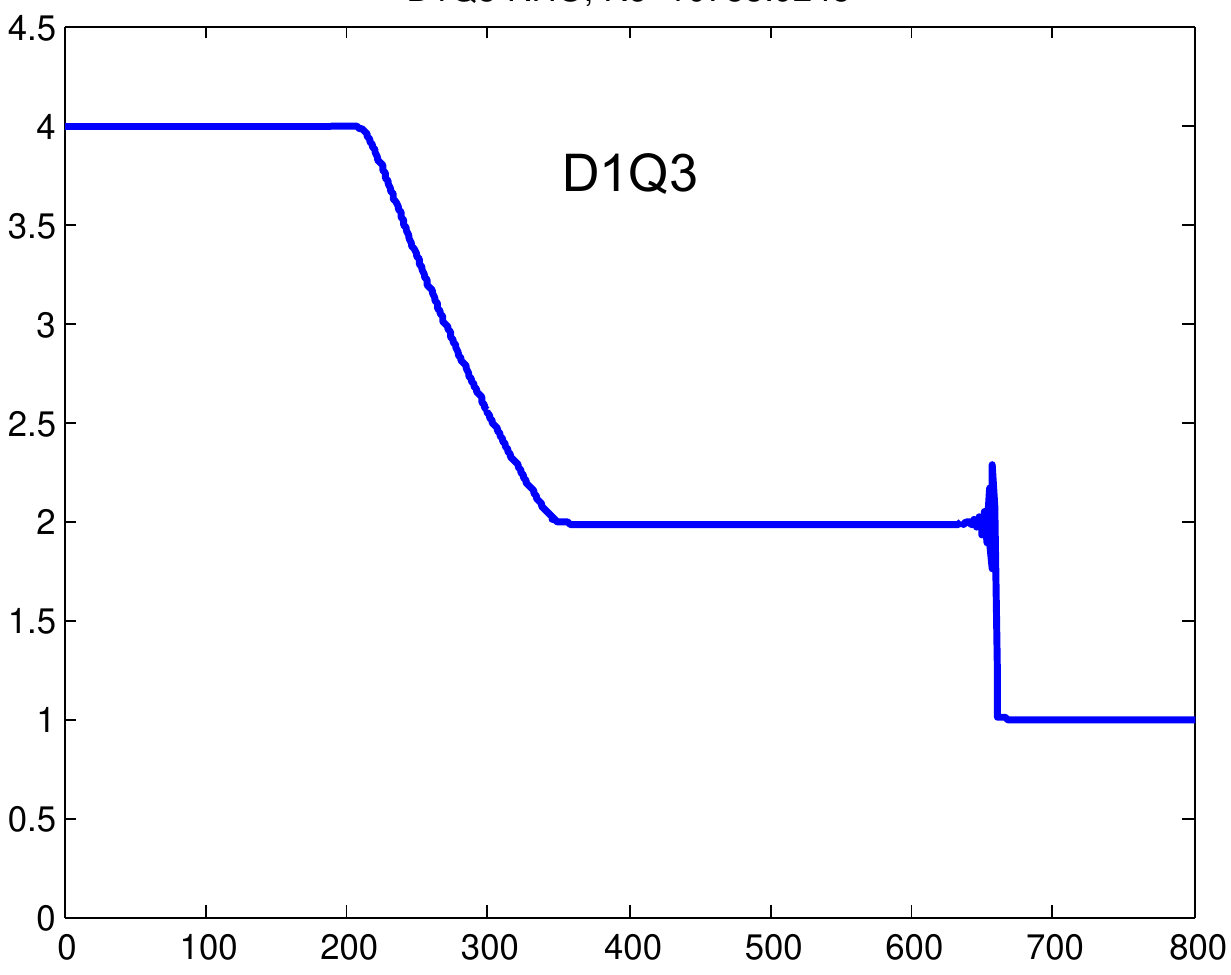}
        \includegraphics[width=0.45\textwidth]{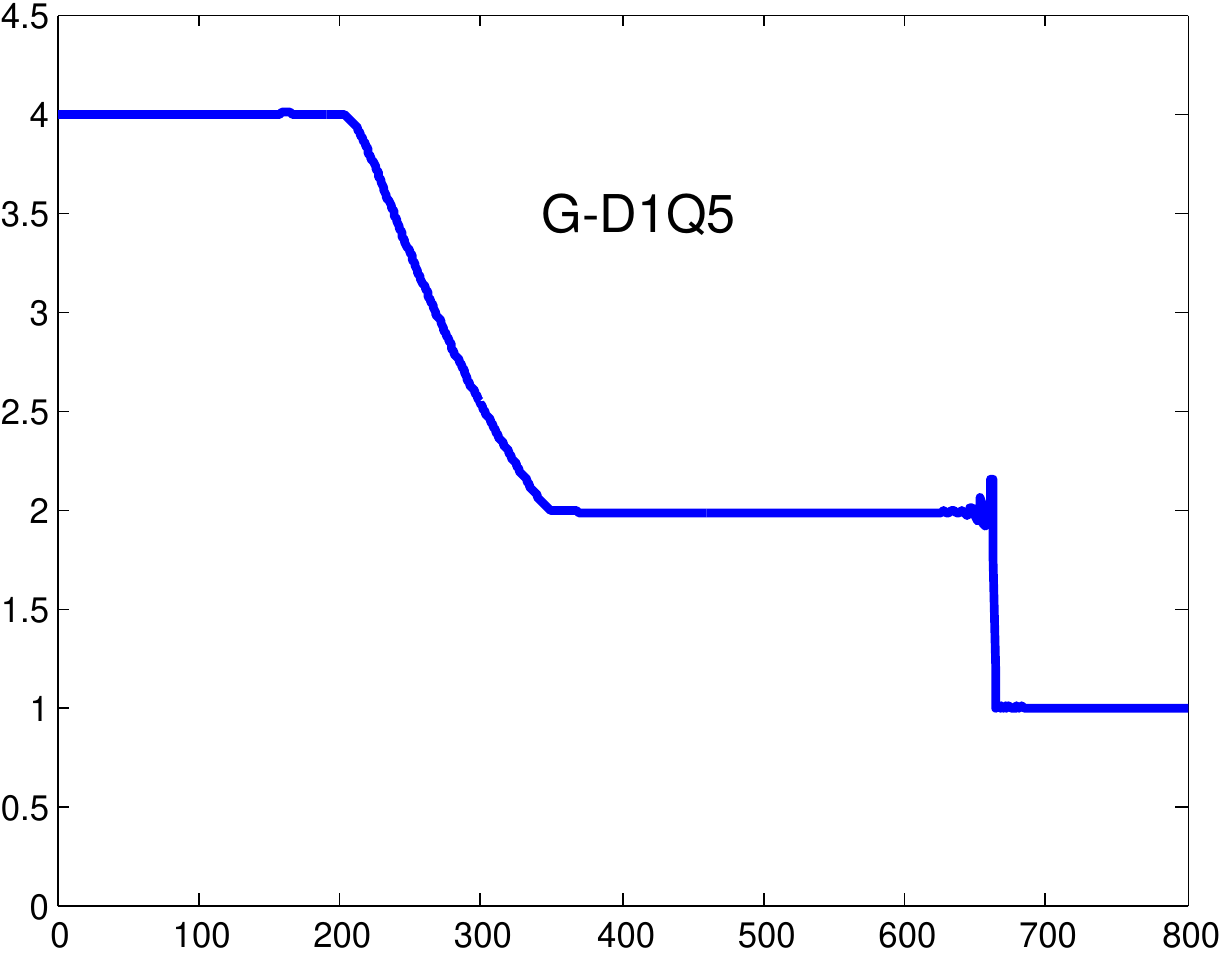}
\end{minipage}
 \begin{minipage}[t]{.99\textwidth}
       \includegraphics[width=0.45\textwidth]{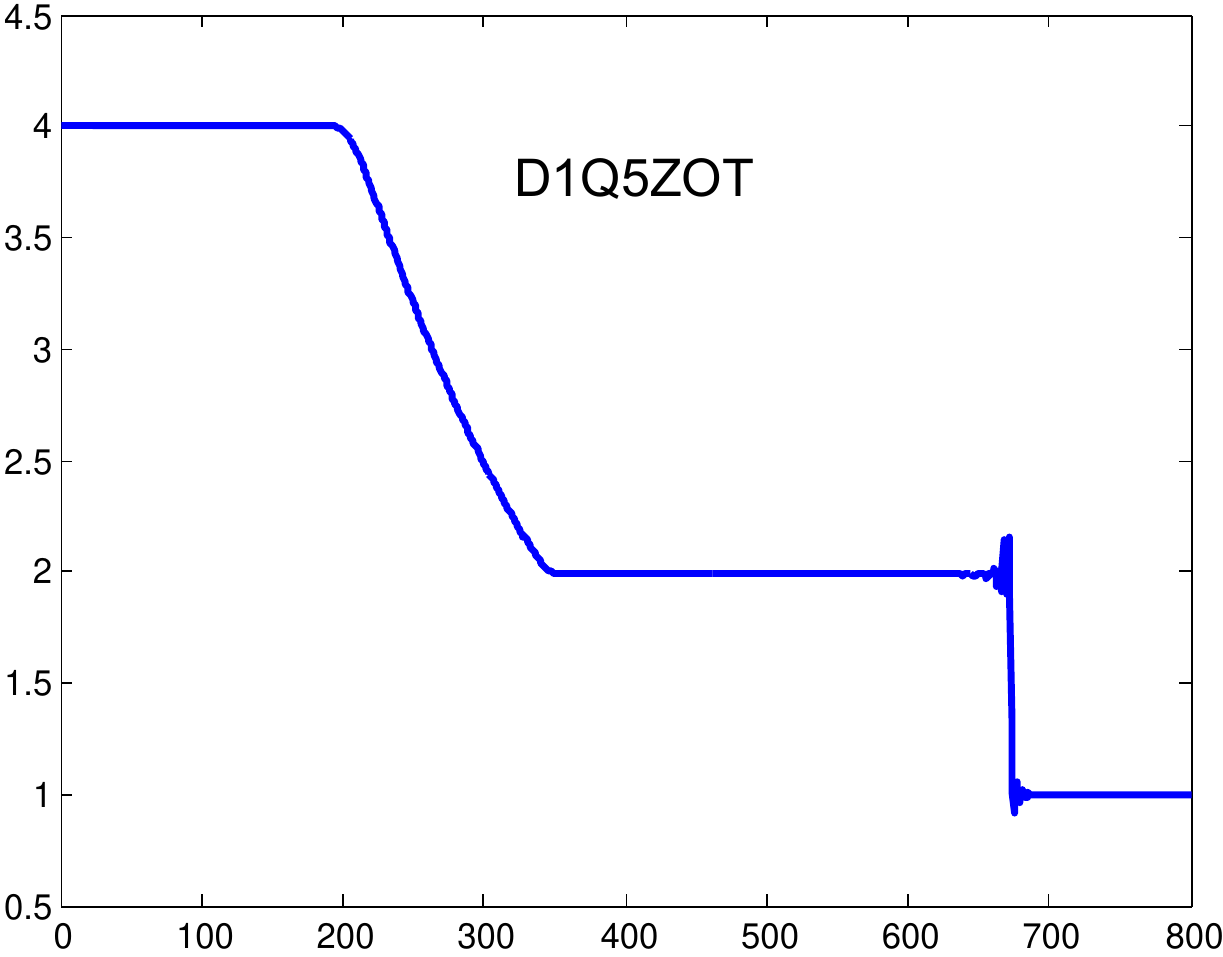}
       \includegraphics[width=0.45\textwidth]{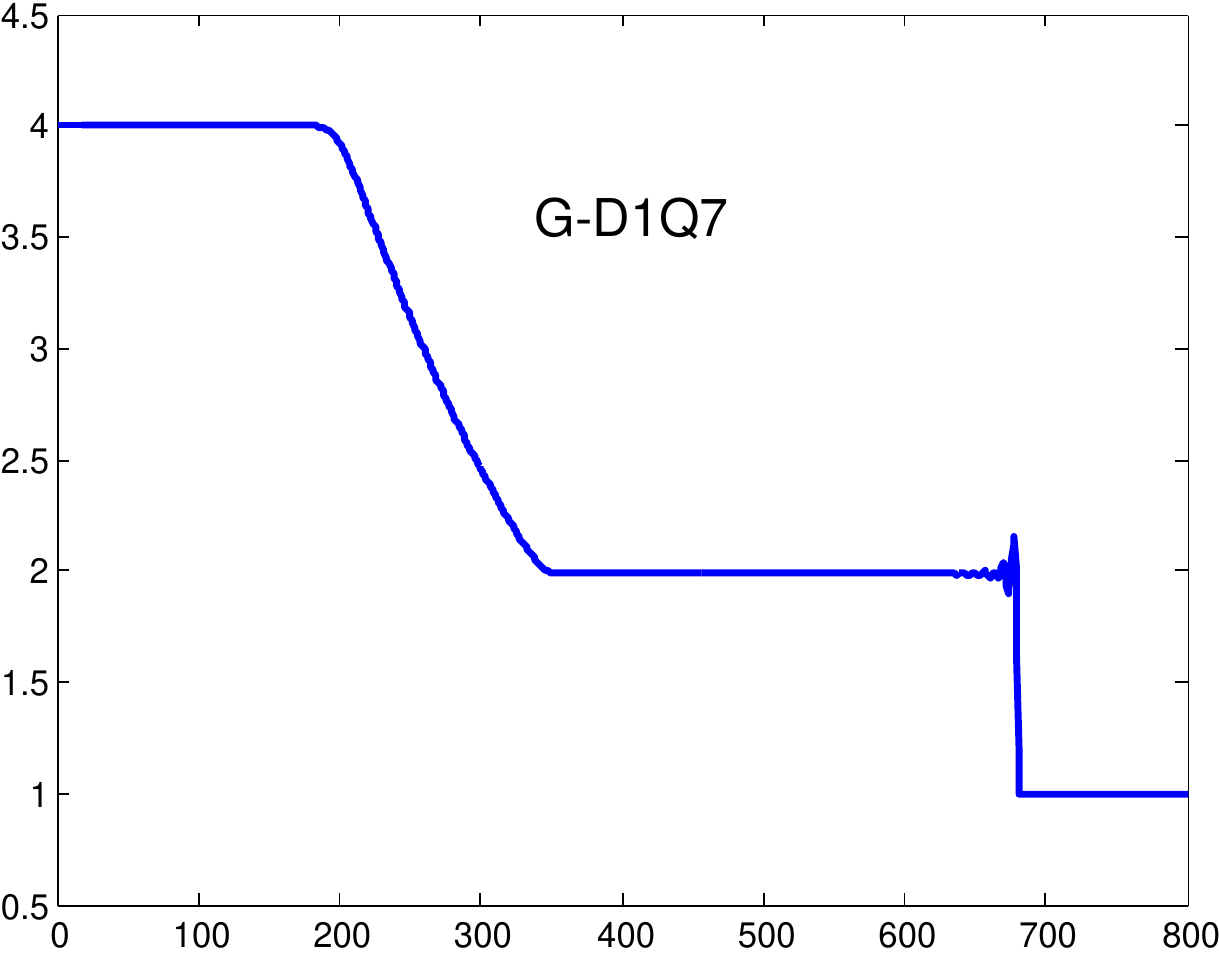}
\end{minipage}
   \caption{ Sod shock tube problem for the viscosity $\nu=0.1$. All the models are calibrated at unit temperature. The initial density ratio is $4:1$, the number of spatial points  equals $800$.}
\end{figure*}

\begin{figure*}
\begin{minipage}[t]{.99\textwidth}
       \includegraphics[width=0.45\textwidth]{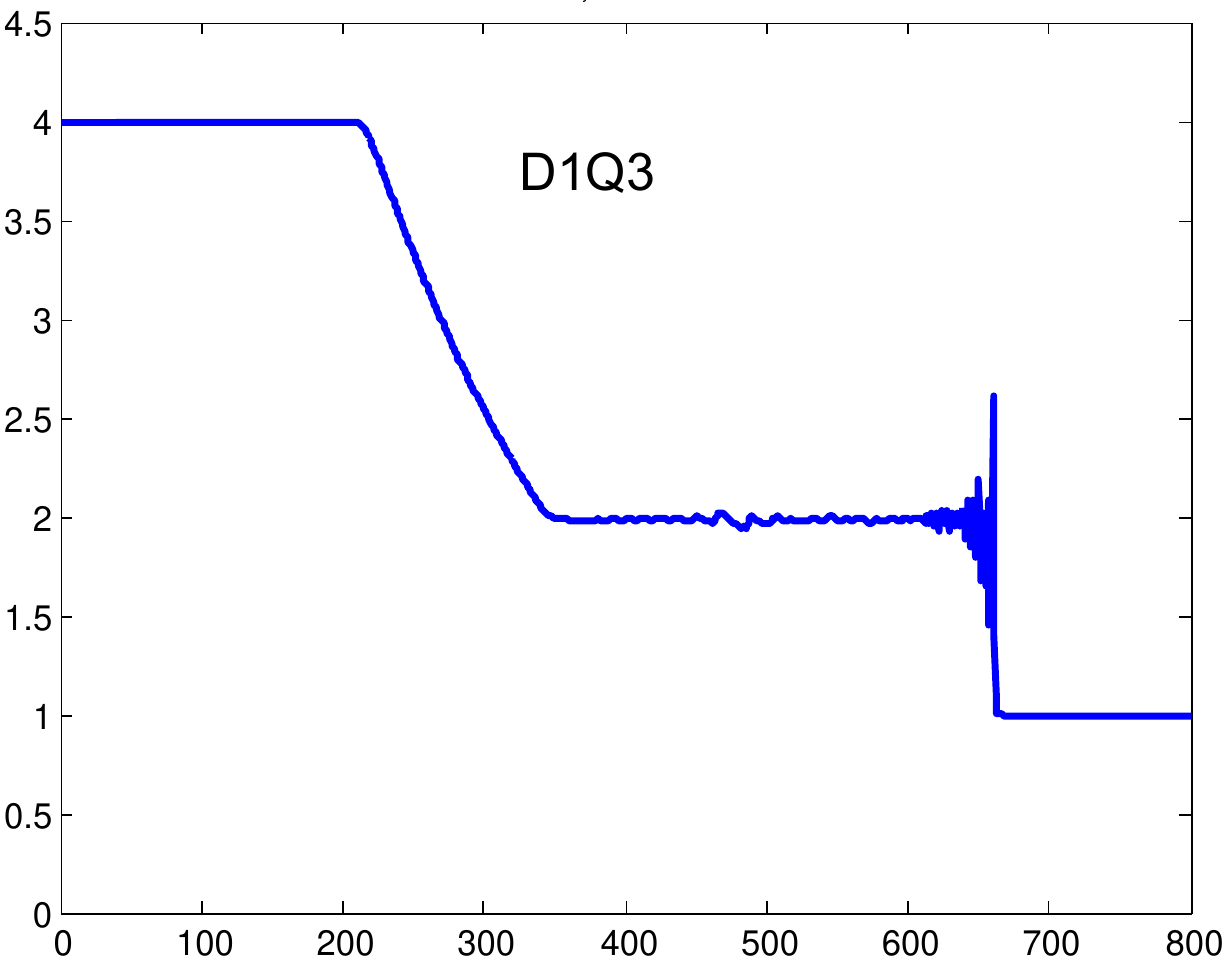}
       \includegraphics[width=0.45\textwidth]{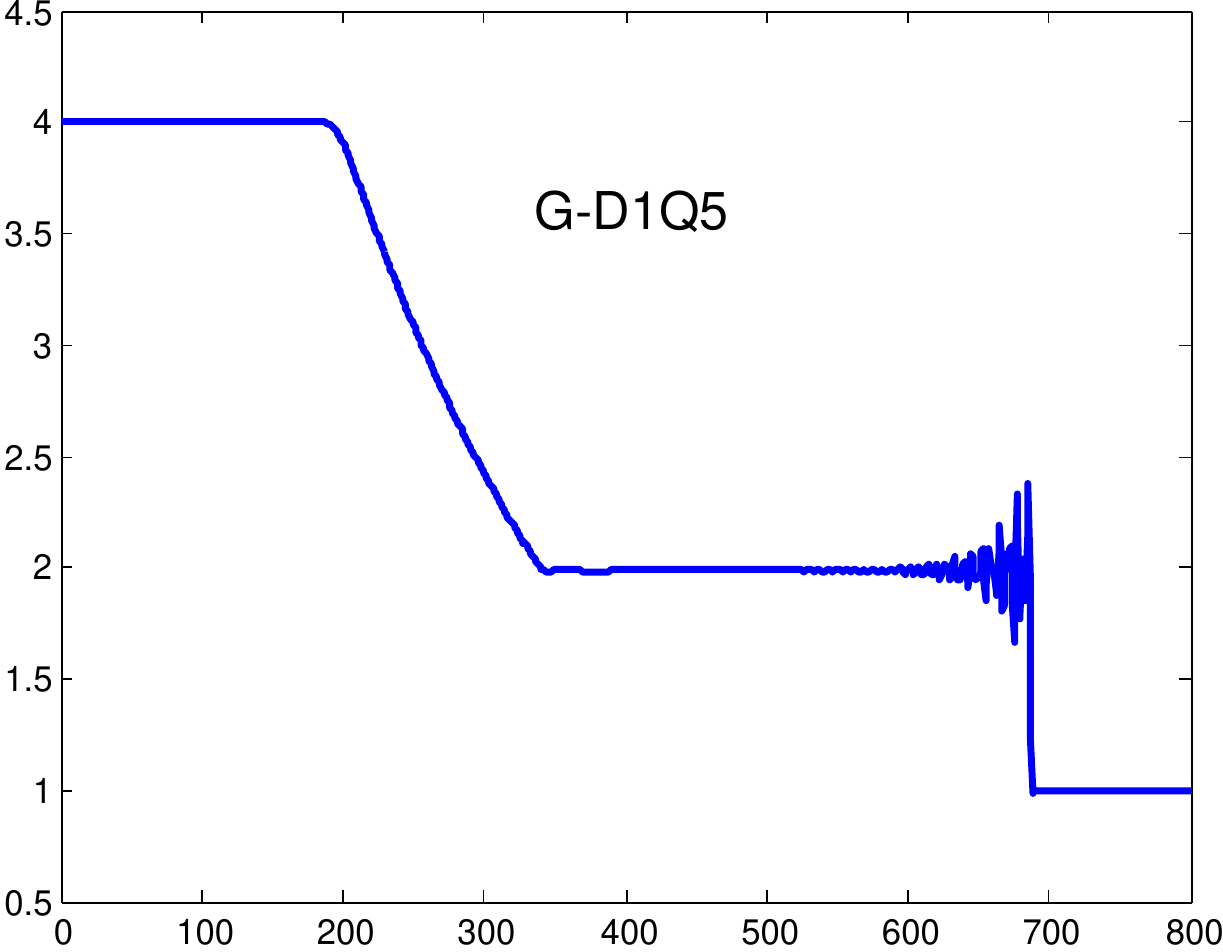}
\end{minipage}
\begin{minipage}[t]{.99\textwidth}
       \includegraphics[width=0.45\textwidth]{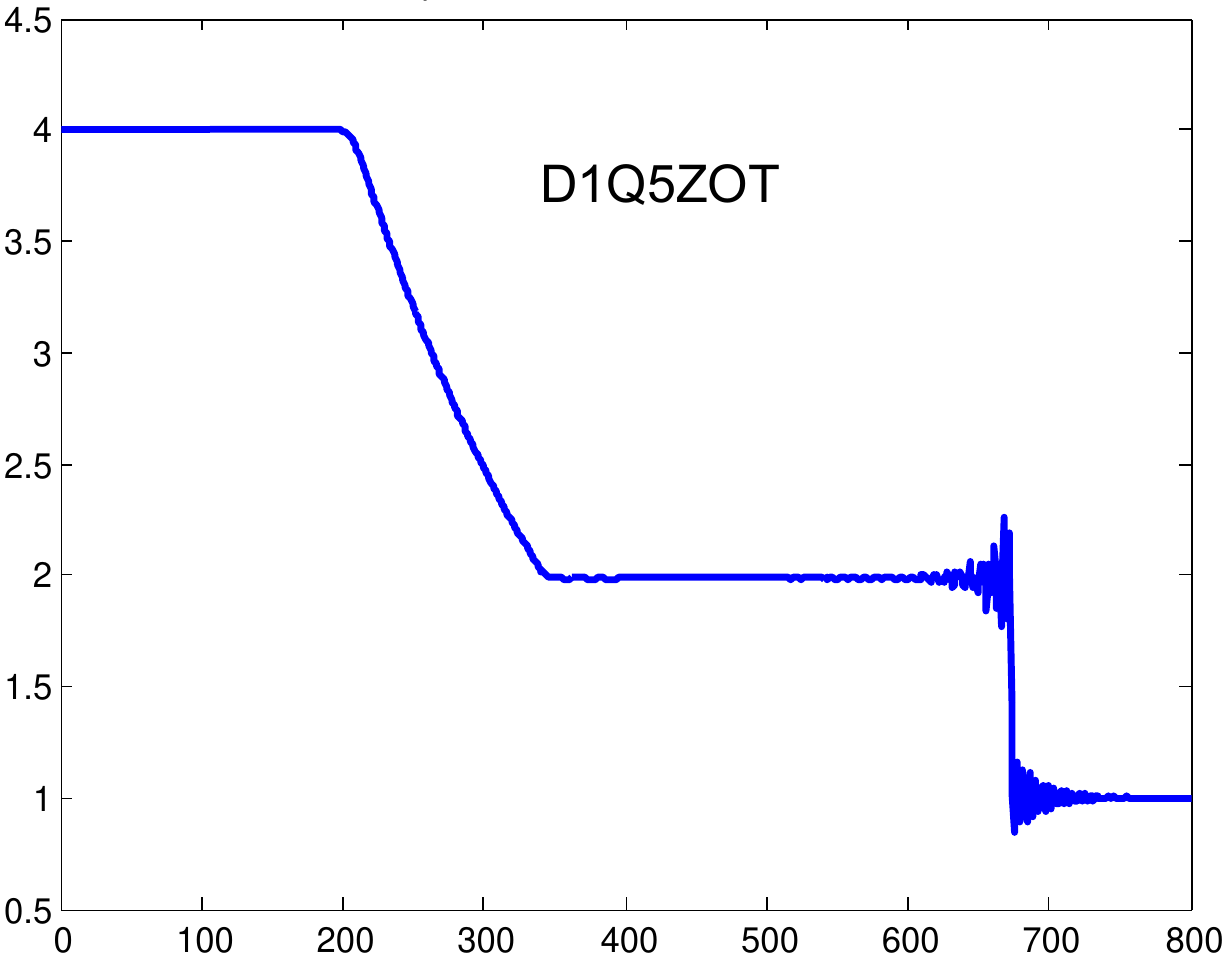}
       \includegraphics[width=0.45\textwidth]{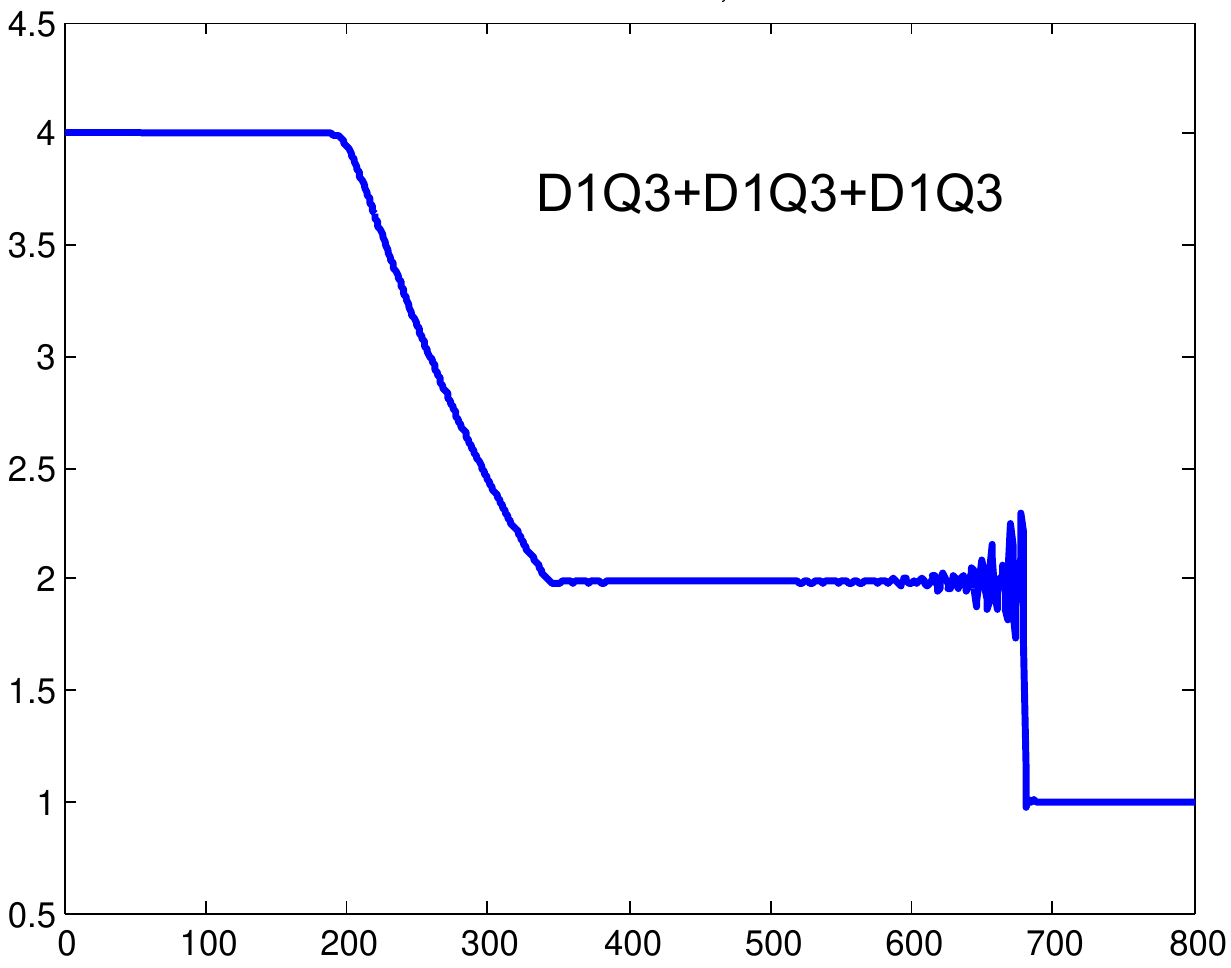}
\end{minipage}
   \caption{ Sod shock tube problem for the viscosity $\nu=0.03$. All the models are calibrated at unit temperature. The initial density ratio is $4:1$, the number of spatial points  equals $800$.}
\end{figure*}

\begin{figure}[h]
  \centerline{\includegraphics[width=0.75\textwidth]{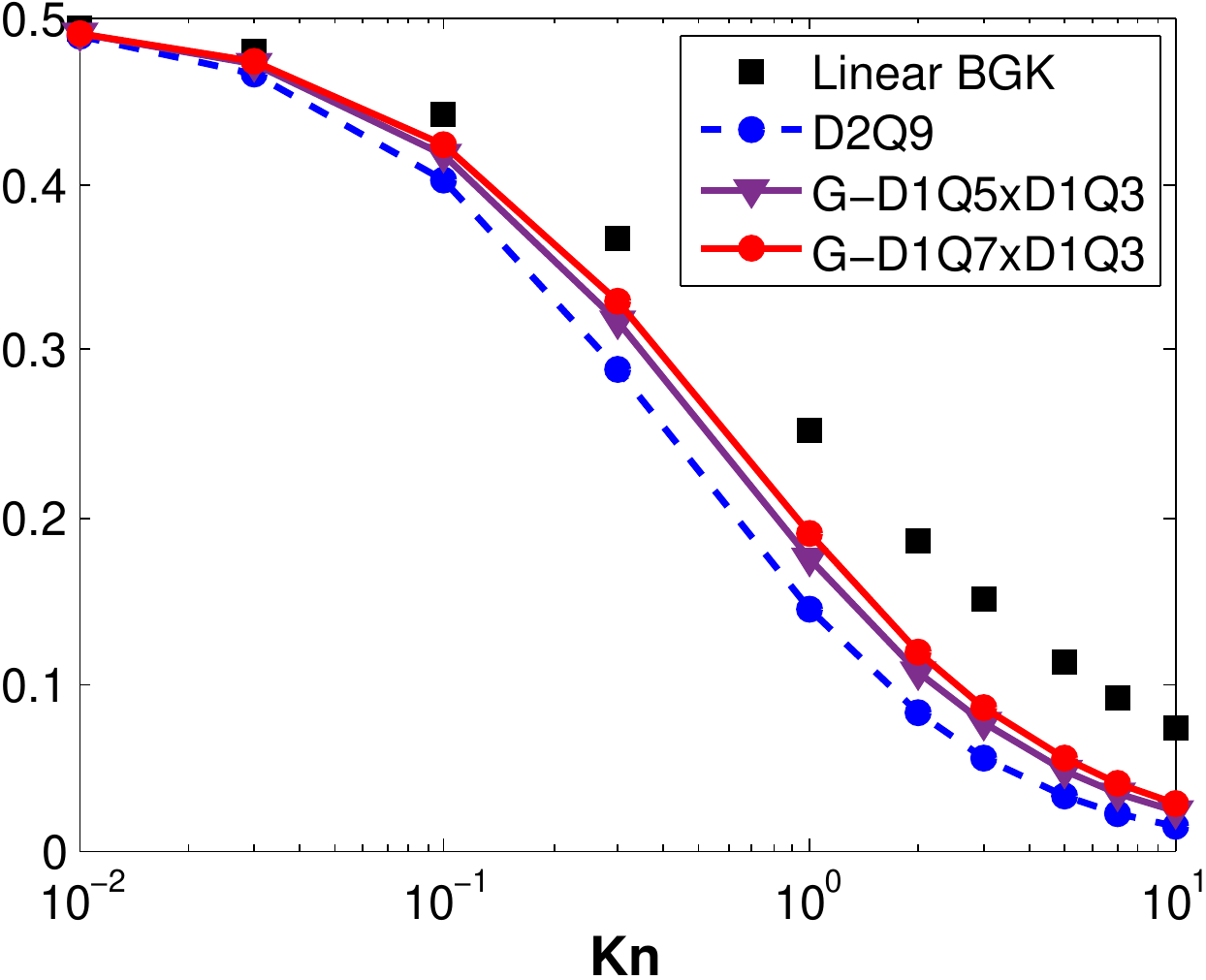}}
  \caption{The slip velocities for the plane  Couette flow.The  benchmark solution (black boxes) was taken from \cite{2015li, 2016jiang} (the numerical solution to the linearized BGK equation).
  }
  \end{figure}
  
\begin{figure}[h]
  \centerline{\includegraphics[width=0.75\textwidth]{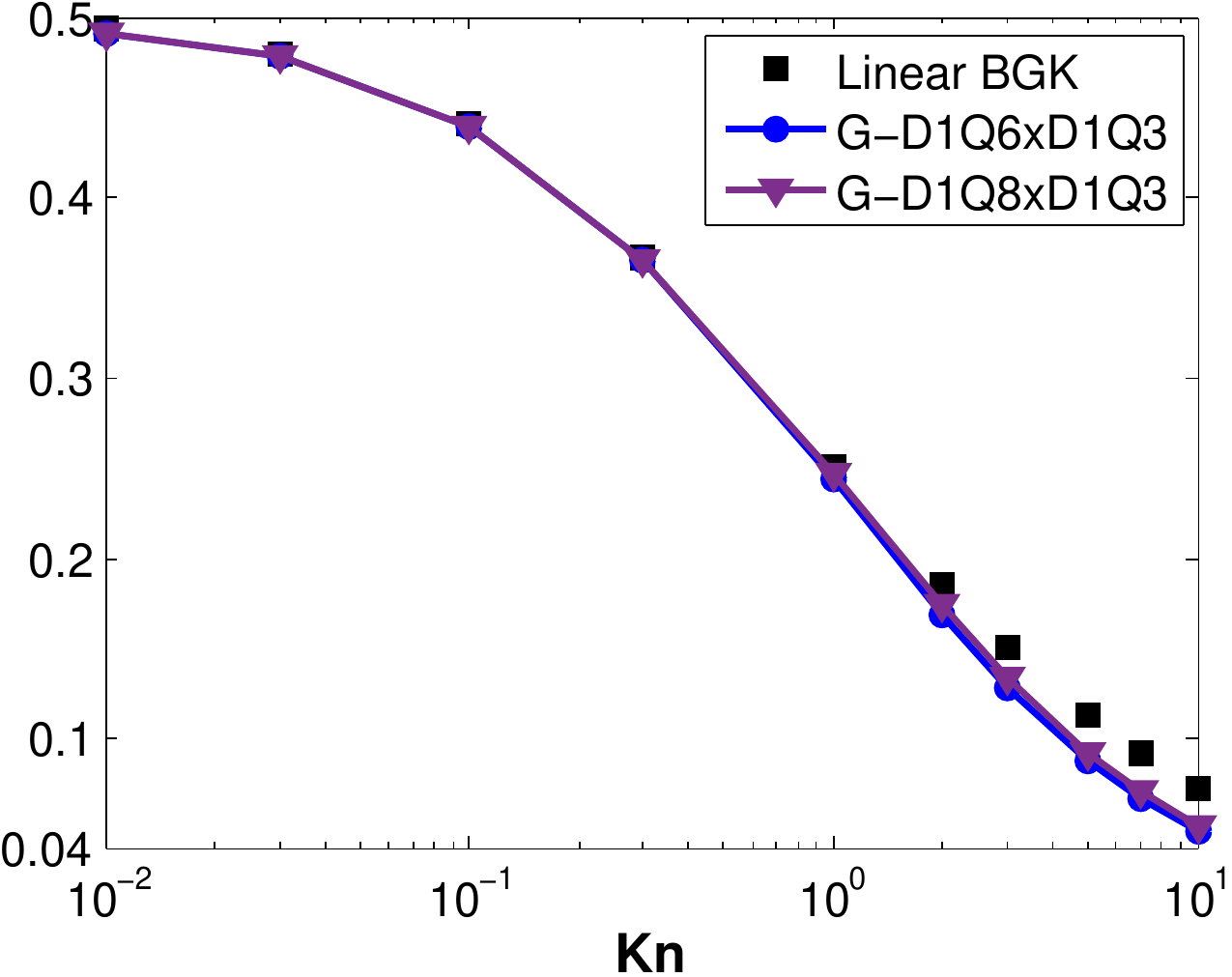}}
  \caption{The slip velocities for the plane Couette flow. The  benchmark solution (black boxes) was taken from \cite{2015li, 2016jiang} (the numerical solution to the linearized BGK equation). 
  }
  \end{figure}
\begin{figure}[h]
  \centerline{\includegraphics[width=0.75\textwidth]{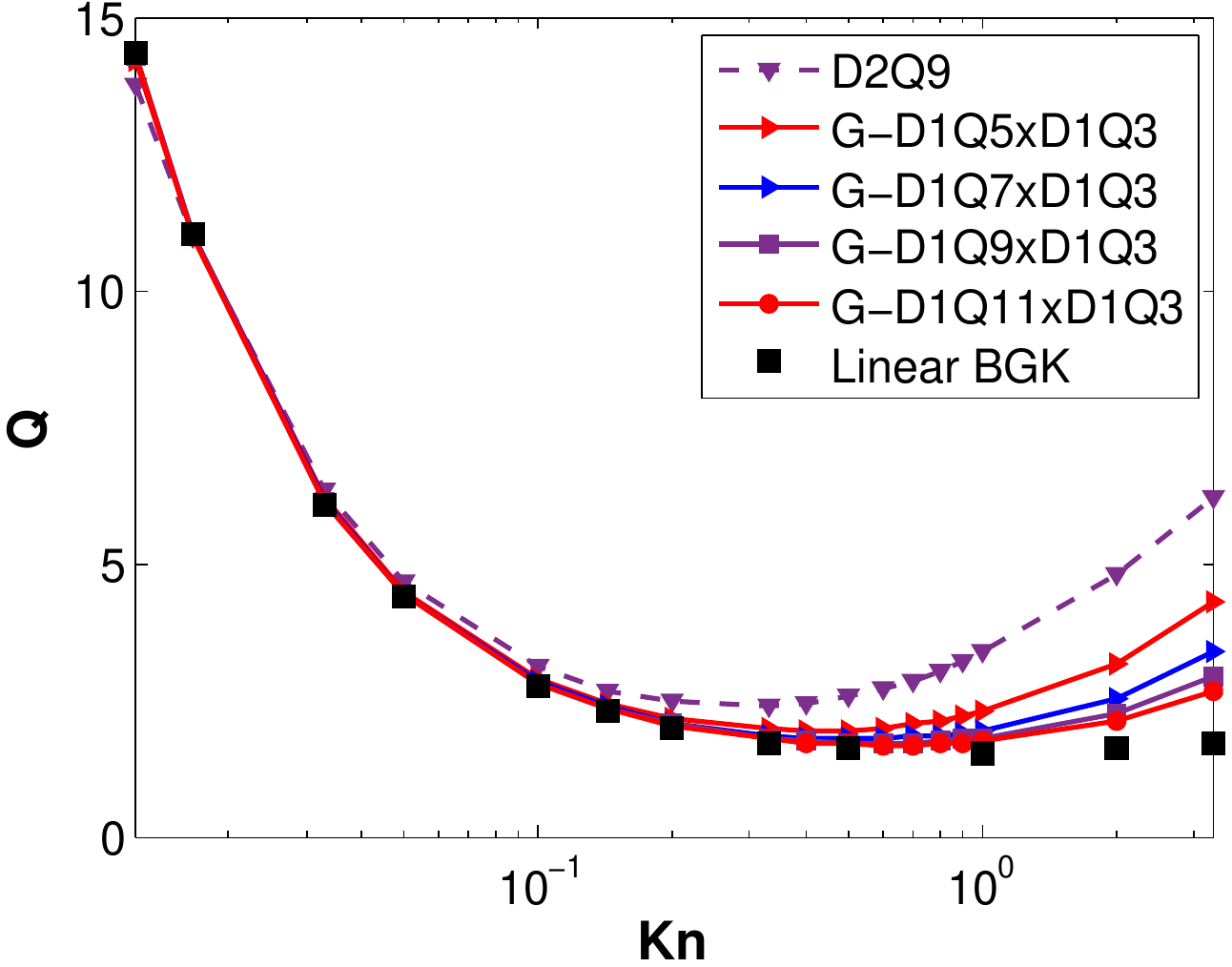}}
  \caption{The Poiseuille  flow across flat walls and the Knudsen paradox. The volumetric flow vs Knudsen number is  presented.
  The volumetric flow rate results for the  linearized BGK solution (defined as Linear BGK in the plot) are taken from \cite{2004cerc}
  }
  \end{figure}
  
  \begin{figure}[h]
  \centerline{\includegraphics[width=0.75\textwidth]{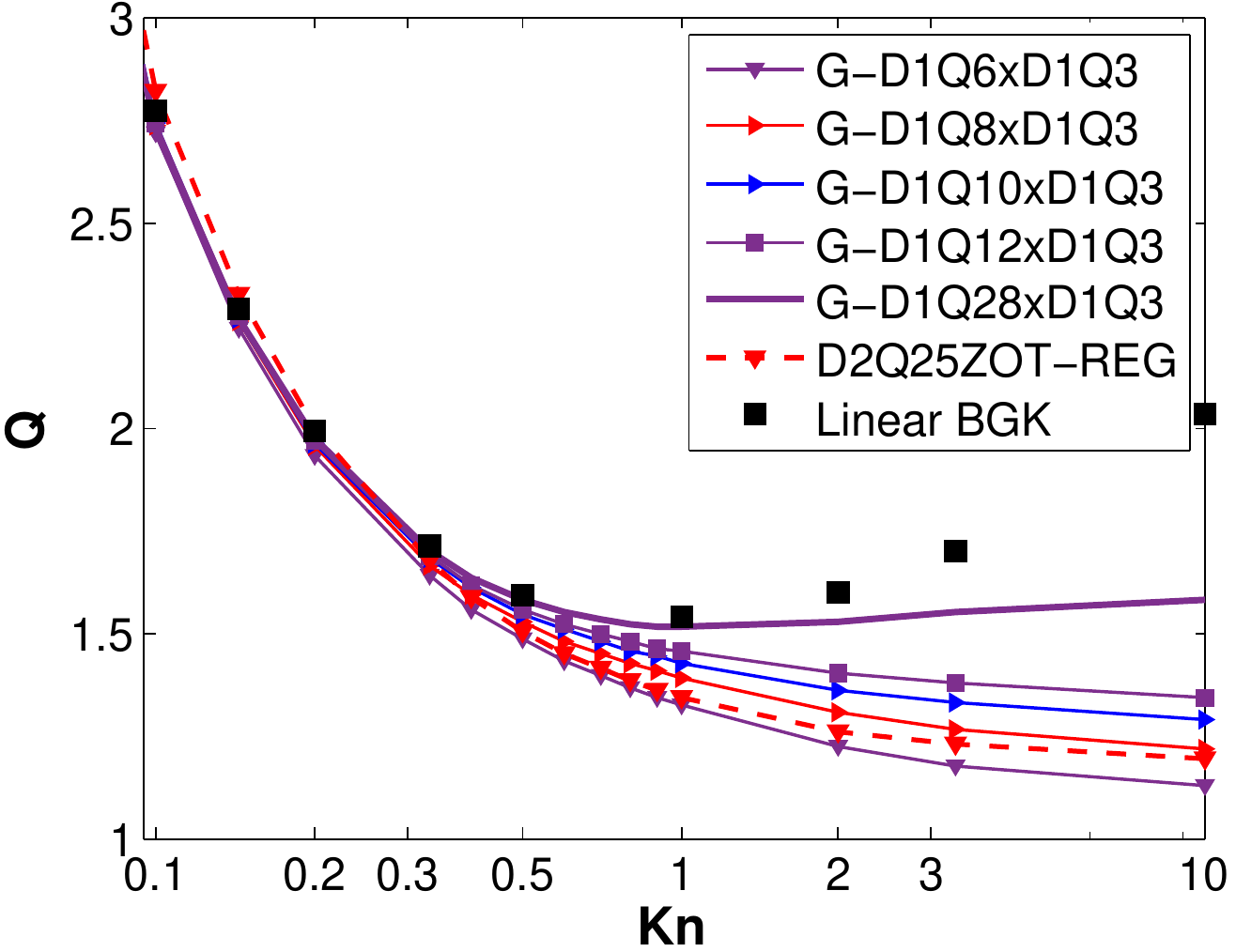}}
  \caption{The Poiseuille  flow across flat walls and the Knudsen paradox for the Knudsen numbers beyond the slip regime ($Kn \geq 0.1$). The volumetric flow vs Knudsen number  is  presented. For the sake of clarity  only transitional and ballistic regimes are shown, $Kn \geq 0.1$.
  The volumetric flow rate results for the  linearized BGK solution (defined as Linear BGK in the plot) are taken from \cite{2004cerc}
  }
\end{figure}

 \begin{figure}[h]
  \centerline{\includegraphics[width=0.75\textwidth, ]{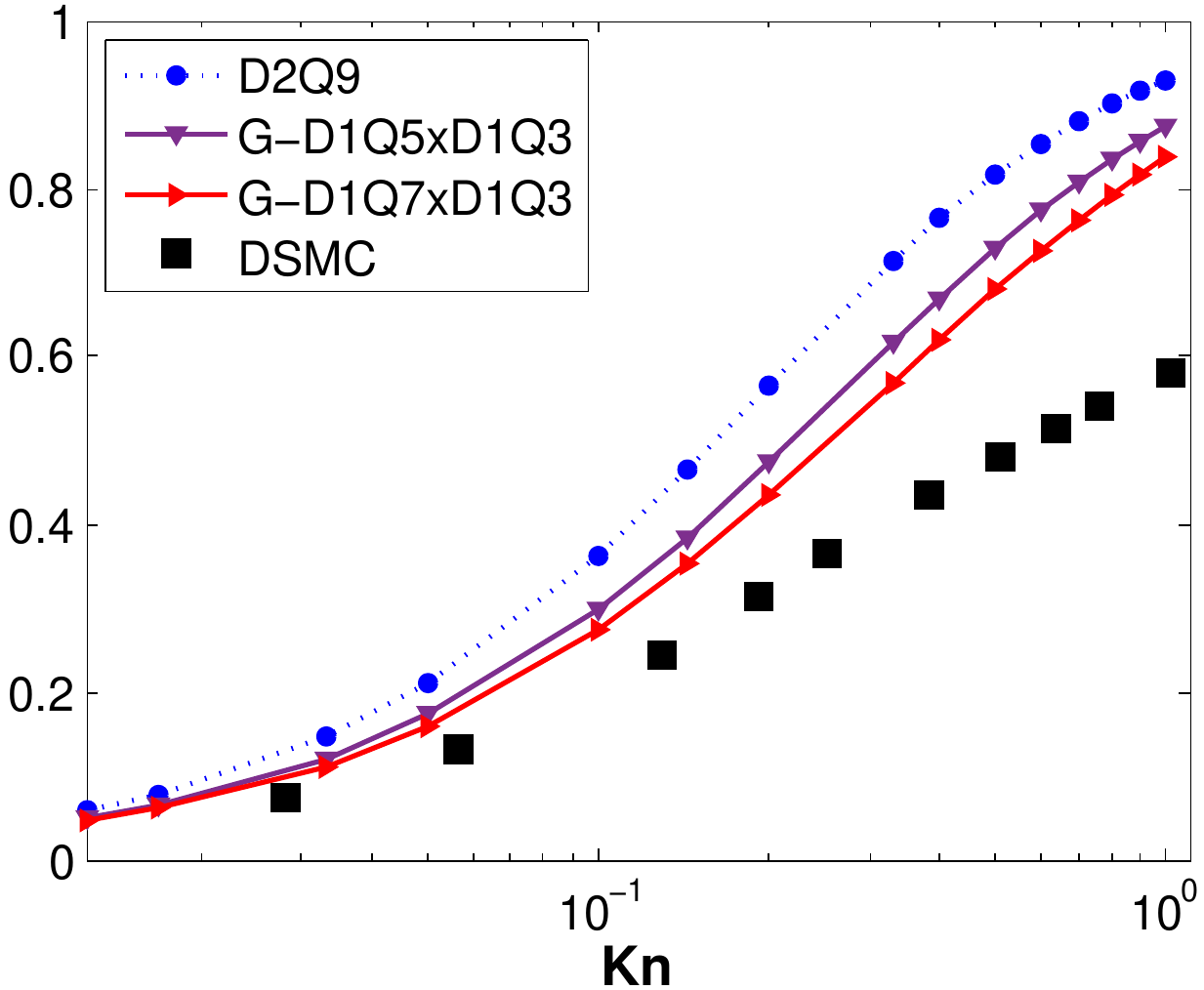}}
  \caption{The slip velocities for the Poiseuille  flow across flat walls at different Knudsen numbers (the  slip velocities are defined as $u(0)/u(H/2)$, where $u(0)$ is the velocity at the wall). The benchmark slip velocities (DSMC) are taken from \cite{2016feucht_schleif}.
  }
\end{figure}

\begin{figure}[h]
  \centerline{\includegraphics[width=0.75\textwidth, ]{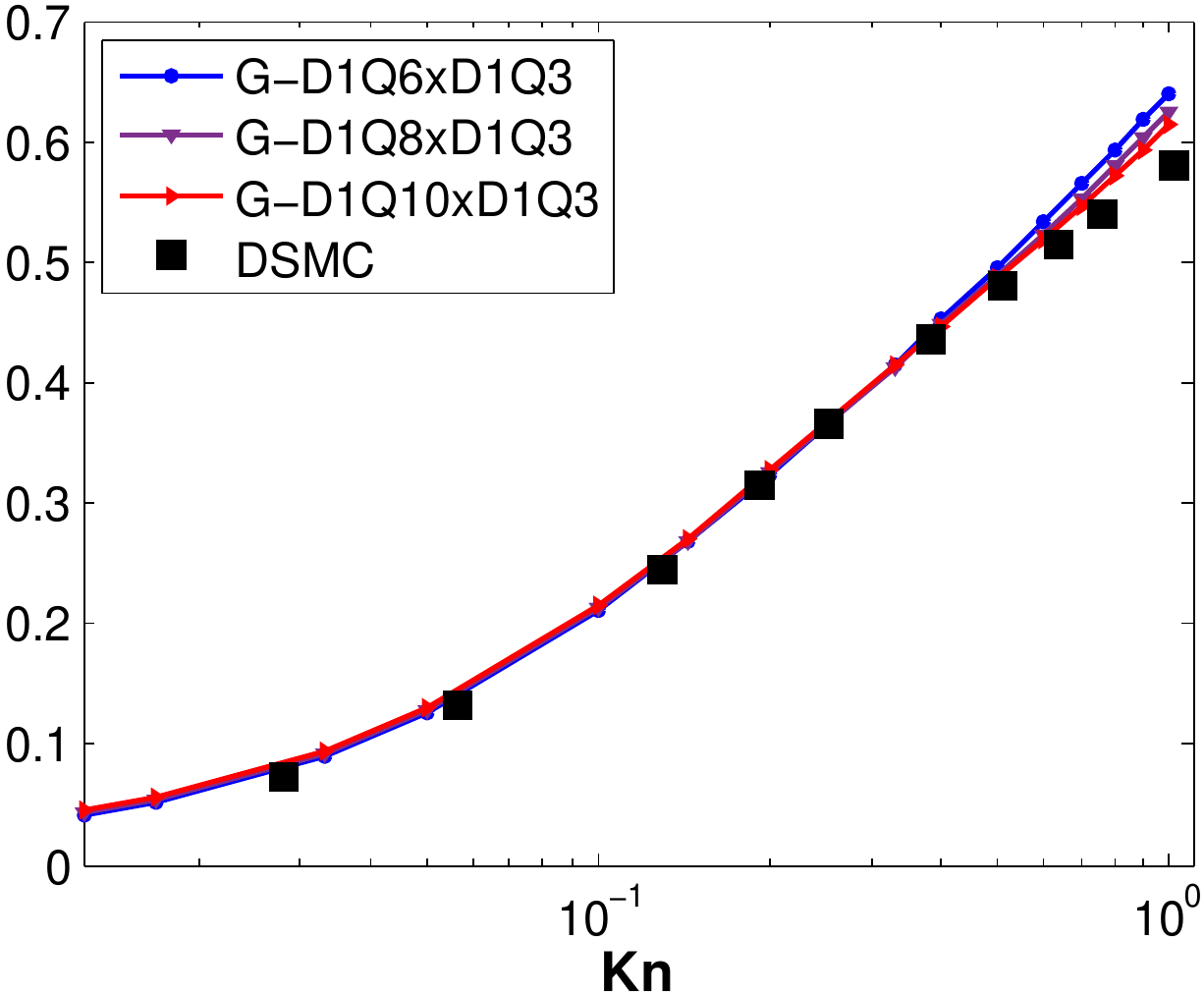}}
  \caption{The slip velocities for the Poiseuille  flow across flat walls at different Knudsen numbers (the  slip velocities are defined as $u(0)/u(H/2)$, where $u(0)$ is the velocity at the wall). The benchmark slip velocities (DSMC) are taken from \cite{2016feucht_schleif}.
  }
\end{figure}

\begin{figure}
  \begin{minipage}[t]{.99\textwidth} 
       \includegraphics[width=0.45\textwidth]{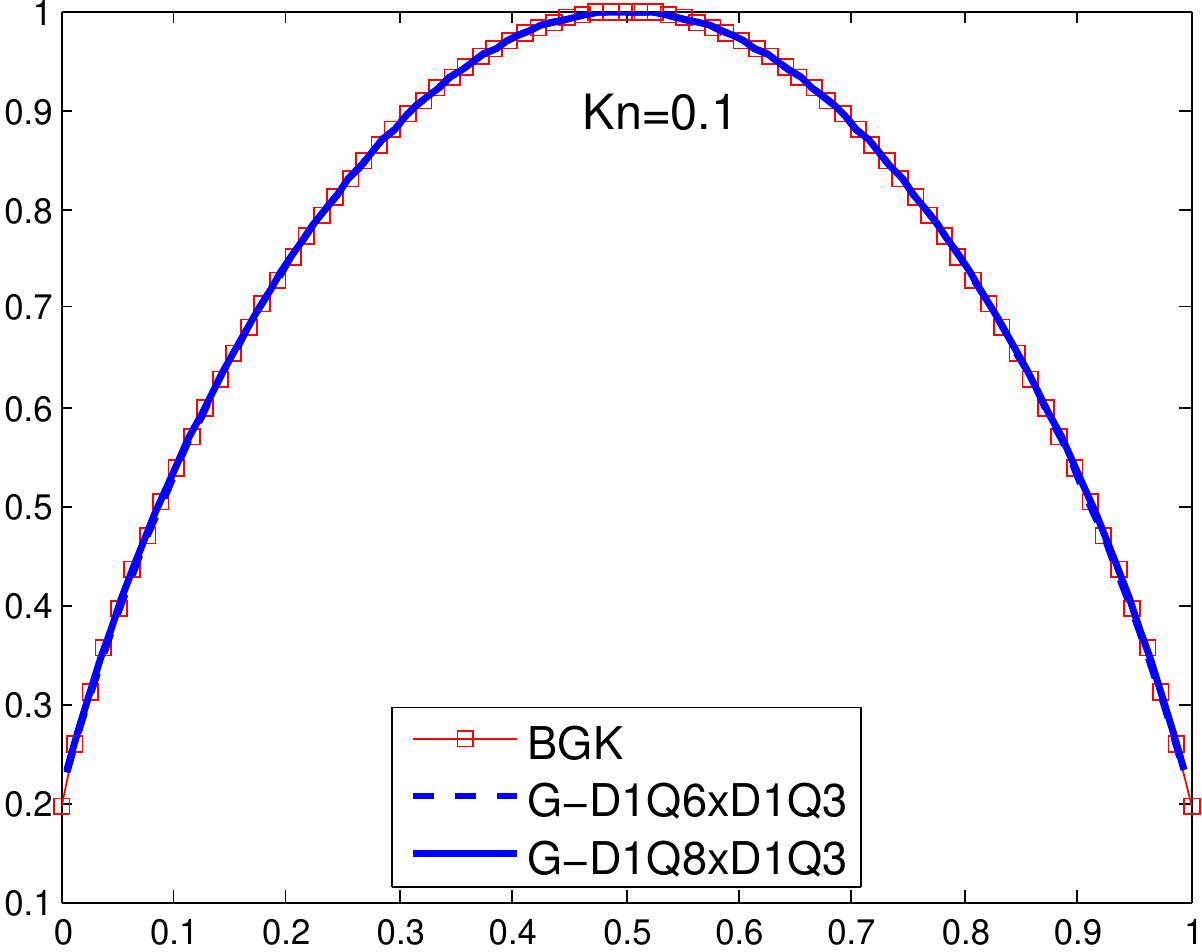}
       \includegraphics[width=0.45\textwidth]{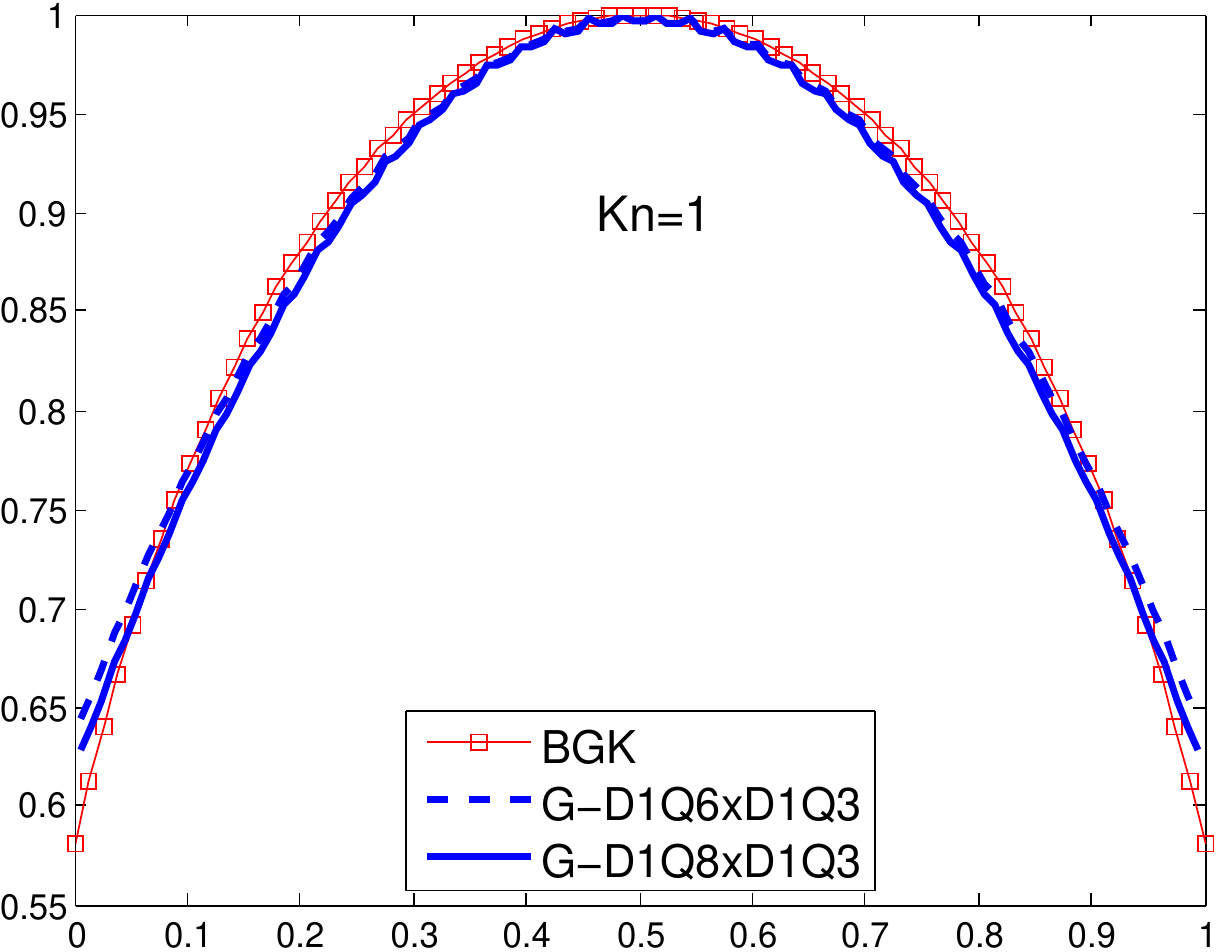}
   \end{minipage}
   \caption{ Normalized velocity profiles for the Poiseuille flow for two Knudsen numbers. The benchmark profile for BGK model (denoted as BGK in the plots) was obtained  using upwind  difference scheme.}
\end{figure}

\section{Test problems: the Sod shock tube, Couette flow, Poiseuille flow }
\subsection{Sod shock tube}
The  first test case  is  1D Sod shock tube  problem. The  initial condition is a step density profile:  $\rho=4, x \leq H/2$ and $\rho=1, x>H/2$,
$H$ is the  length of the domain,  $800$ spatial nodes were used.

As a benchmark   $D1Q5ZOT$ LB model was  adopted, this model  has  good  stability \cite{2006chikatamarla,2009chikatamarla}. The models were calibrated at the unit temperature $\theta=1$. 
All the models show good stability for moderate viscosity $\nu \sim 0.1$ ($\nu=\theta \tau$) but show some overrelaxation effects (oscillations) (Fig. 1). When the viscosity was decreased to $0.03$
(this  corresponds to the Reynolds  number close  to $30-40\times 10^3$, where the Reynolds  number  is defined as $uH/\nu$)
the  oscillations are  amplified (Fig. 2). The  overall oscillation magnitudes are largest for $D1Q3$, the further small decrease in viscosity leads to breakdown of the solution to $D1Q3$ model while all the  other models are able to reproduce  the solution (though oscillations are growing). 
 As a result, one can  conclude that the models in the hierarchy for $N>1$ have  better stability than the conventional $D1Q3$.  

Interestingly, that five velocity  models for $(-2c,-c,0,c,2c)$ lattice are supposed to be  unstable \cite{2006chikatamarla,2009chikatamarla}. This  result seems to contradict to the presented Sod shock tube simulations since it was shown that the $G$-$D1Q5$ model from the $2N+1, N=2$ hierarchy has good  stability. This seeming  contradiction has  the following explanation. The result in the papers \cite{2006chikatamarla,2009chikatamarla} is related to the  high order LB model (which exactly reproduces the third order  moments  of the Maxwell distribution) while in the  present case  $G$-$D1Q5$ has  the same order  of isotropy as $D1Q3$ model. 

\subsection{Knudsen number}
For the next two test problems (Couette and  Poiseuille flow)  the rarefaction measure or Knudsen number should be introduced. The  modeling  results will be compared with the data
from literature in which numerous definitions are used. Thus this definition should be considered in detail for consistency. 

Following the  paper \cite{2008kim} the viscosity based Knudsen number  is  introduced 
$$
k=\frac{\lambda}{H}=\sqrt{\frac{\pi}{2\theta}}\frac{\nu}{H},
$$
where $\lambda, H$ are the mean free path and width of the channel for the plane Couette or Poiseuille  problem; $\nu=\theta \tau$ is the viscosity,  $\tau, \theta$ are the relaxation time and temperature.

In several papers \cite{1990fukui, 1998sharipov,2004cerc} the  results for the Poiseuille flow   are presented against a rarefaction parameter $\delta$  which is  proportional  to the  inverse of the Knudsen number 
$$
\delta=\frac{\sqrt{\pi}}{2k}  ,
$$
therefore another  definition of the  Knudsen number  can be introduced
\begin{equation}\label{kn}
Kn=\delta^{-1}=\sqrt{\frac{2}{\theta}}\frac{\nu}{ H}
\end{equation}
In the  present  paper I will stick to the definition (\ref{kn}) which is most convenient since the benchmark results for the Poiseuille and Couette
flows are  presented for $\delta$ or $Kn$ \cite{1990fukui,1998sharipov,2004cerc,2015li,2016jiang}.

\subsection{Couette flow}
The plane 2D Couette flow  is considered.  For this  flow the parallel plates  move in opposite  direction with the velocities $\pm U_w$ respectively. The  magnitudes of the  velocities are taken small, such that the flow velocity is $Ma \sim 10^{-4}$.
The  kinetic boundary conditions for high-order lattices are stated at the walls \cite{2014meng, 2002ansumali}, $200$ spatial nodes  between the walls are used in the computations.

The slip velocities are defined as $u(0)/(2U_w)$, where $u(0)$ is the velocity at the wall for the LB models. The slip  velocities  are  presented in Fig. 3 and Fig. 4  for the Knudsen numbers  varying from $10^{-2}$ to $10^{1}$. The  high-precision solutions to the  linearized BGK equations are chosen as  benchmark \cite{2015li, 2016jiang}.

It is worth to mention that for 2D models  in the form $G$-$D1Q5 \times D1Q3$, $G$-$D1Q7\times D1Q3$ or $G$-$D1Q6 \times D1Q3$, $G$-$D1Q8 \times D1Q3$ and etc, the  parts $G$-$D1Q5$, $G$-$D1Q6$, $G$-$D1Q7$ and etc are responsible for the dynamics transverse to the flow direction (perpendicular to the walls) while the component $D1Q3$ is responsible to the streamwise direction. 

Obviously $D2Q9$ fails to reproduce Knudsen layer and understates the slip velocities for $Kn>0.05$. This behavior is well-known \cite{2007ansumali, 2008tang} and  is  predicted analytically \cite{2007ansumali}.
The results for the models $G$-$D1Q5 \times D1Q3$, $G$-$D1Q7 \times D1Q3$  based  on the summation procedure presented earlier (\ref{maineq01})-(\ref{maineq05}) are significantly better than $D2Q9$ for all Knudsen numbers. Nevertheless, in comparison with  the results for the fourth-order off-lattice  $D2Q16$  model (Fig. 1 in the paper \cite{2007ansumali}) the models $G$-$D1Q5 \times D1Q3$, $G$-$D1Q7 \times D1Q3$ show  worse precision. The  model  $D2Q16$ predicts Knudsen layer at  least qualitatively. Also the  lattice velocities for $D2Q16$ model do not  have components  parallel to the walls.
The latter seems to be even more important than the fact that $D2Q16$ is  high order LB model. The  results  for $G$-$D1Q6 \times D1Q3$, $G$-$D1Q8 \times D1Q3$ models
(no lattice velocities parallel to the walls) support  this  idea. The precision for these models is much better than $G$-$D1Q5 \times D1Q3$, $G$-$D1Q7\times D1Q3$ and also $D2Q16$  for all Knudsen numbers (Fig. 4).

 The  positive  effect of zero velocity removal is  thoroughly 
 explained in \cite{2011meng,2016feucht_schleif}.  Zero velocity usually has  the lattice weight significantly greater than the weights of the other velocities, on the other  hand  zero-velocity weight does not influence  half-moments (or half-fluxes) which enter the kinetic boundary conditions. Therefore, the  wall half-moments are underestimated. The models $G$-$D1Q6 \times D1Q3$, $G$-$D1Q8 \times D1Q3$ from $(2N+2)$ hierarchy (\ref{maineq01_noball})-(\ref{maineq04_noball}) do not have  wall-parallel velocities  moreover the form of their local equilibrium is close  to Gaussian which result  in a good  reproduction of the kinetic boundary conditions. 
 
\subsection{Poiseuille  flow}
The force driven 2D Poiseuille flow is considered. The kinetic boundary conditions are stated at two  parallel walls \cite{2014meng, 2002ansumali}.
Similarly to the  previous case  the  parts $G$-$D1Q5$, $G$-$D1Q6$, $G$-$D1Q7$ for $G$-$D1Q5 \times D1Q3$, $G$-$D1Q6 \times D1Q3$, $G$-$D1Q7 \times D1Q3$ and etc are responsible for the dynamics transverse to the flow direction (perpendicular to the walls).
The force  is taken in the  linearized form \cite{2017kruger}
$$
F_{ij}=W_iw_jc_jF, 
$$
where $ i=1 \ldots 2N+1$ or $ i=1 \ldots 2N+2$ and $j=1,2,3$; $F$ is the  force amplitude, $W_i$ are the weights  for the
(the values  of local equilibrium state taken at $\rho=1, u=0$, see example in Appendix A) for the lattices $G$-$D1Q5$, $G$-$D1Q6$, $G$-$D1Q7$ and etc and $w_j, c_j$ are the weights and  the lattice velocities for $D1Q3$ model (calibrated at  unit temperature): $(1/6, 4/6, 1/6)$ and $(-\sqrt{3},0, \sqrt{3})$ respectively; 
$200$ cross-stream spatial nodes were used in the computations.
The amplitude of the  force $F$ is taken small such that the flow velocity is  of order $Ma^{-4}$, moreover instead of using the force term the boundary conditions  with pressure variations were tested \cite{2007kim}, they give  very similar  results  to the force-driven case.  

The reproduction of the  Knudsen paradox i.e. the shape  of the the volumetric flow  with a minimum near $Kn=1$ is  challenging  for the Lattice Boltzmann method.  The  volumetric flow is defined as \cite{2008kim}
$$
Q=\frac{\delta}{4 U_0 H} \int_{0}^H u(s) ds,
$$
where $U_0=FH^2/(8\rho \nu)$ is the centerline velocity for the Navier-Stokes equation with no-slip boundary conditions, $H$ is the  distance  between the walls, $\rho$ is the gas  density, $\nu$ is the  viscosity, $u$ is  the streamwise velocity.

The  straightforward application of high-order lattices  is not sufficient  for the reproduction of the rarefied flow effects \cite{2011shi, 2011meng}. For instance, the Knudsen minimum is well reproduced   when extreme high-order  LB models are used \cite{2011meng}. 
There exist numerous  approaches to improve the results for the Poiseuille flow. The  multiple relaxation LB approach is able to capture non-equilibrium rarefaction effects in several test flows \cite{2008guo, 2011li_Micro, 2017su}.
The regularization of LB models significantly improves the results  \cite{2006zhang,2015mont} yet the Knudsen minimum is not obtained. The application of the regularization  with an additional inclusion of two relaxation times (dependent on Knudsen number)
leads to the prediction of the Knudsen minimum \cite{2007niu}. Another solution is  the alternation of even and odd high-order LB schemes and averaging the results \cite{2011izarra}. The high-order on-lattice models
with correct half-fluxes (wall half-moments) \cite{2016feucht_schleif} show good accuracy for $Kn<1$, the thermal off-lattice schemes with exact half-fluxes based on $S$ kinetic model are able to predict the volumetric flow for a wide range of Knudsen numbers \cite{2016ambrus2,2016ambrus, 2014ambrus}.

In the  present paper  un-regularized on-latice LB models with  single relaxation time are studied (except the regularized $D2Q25ZOT$ which is  used as  benchmark). The   results  for the models from $2N+1$ hierarchy (\ref{maineq01})-(\ref{maineq05}) are   presented in Fig.5 and Fig. 7. The  precision is  significantly increased over $D2Q9$.  The  volumetric flow modeling results  in Fig. 5 can serve as an apparent example of the  convergence  for the $2N+1$ hierarchy to the BGK equation.

In ballistic regime runaway effects prevent the correct computation of the flux for $2N+1$ hierarchy (\ref{maineq01})-(\ref{maineq05}) while for the models from $2N+2$ hierarchy (\ref{maineq01_noball})-(\ref{maineq04_noball}) runaway effects are absent but the flow does not  have minimum. This result is very  typical for LB models with a single relaxation time: the models with velocities parallel to the wall suffer from runaway effects, while the  models which are free of such lattice velocities do not reproduce the Knudsen minimum, see \cite{2008kim,2008kim_sols, 2015mont,2016feucht_schleif}. 
Nevertheless, the models from $(2N+2)$ hierarchy (\ref{maineq01_noball})-(\ref{maineq04_noball}) show excellent accuracy in slip and  transitional regimes ($Kn<1$), Fig.6, Fig.8 and Fig. 9.
The  regularized  $D2Q25ZOT$ model (2D analog of the regularized $D3Q41$ model applied in \cite{2015mont}) 
was also implemented for the  comparison with the models from $2N+2$
hierarchy. For clarity the  volumetric  flow is shown in Fig. 6 only for the transitional and ballistic regimes  ($Kn \geq 0.1$).  All the  models  have  good precision in slip regime only  the regularized $D2Q25ZOT$ slightly overestimates the flow. This  at least in  qualitative agreement with the  results  from \cite{2015mont},
where  the slight over prediction of the flow for the regularized  $D3Q41$ is  observed in the slip and   transitional regimes. 
In the part of the transitional regime and  ballistic regime the regularized $D2Q25ZOT$ performs better than $G$-$D1Q6\times D1Q3$, Fig. 6. The  next  models in the  hierarchy $G$-$D1Q8\times D1Q3$, $G$-$D1Q10\times D1Q3$, $G$-$D1Q12\times D1Q3$  surpass both the regularized $D2Q25ZOT$  and $G$-$D1Q6\times D1Q3$ models for all Knudsen numbers and  monotonically converge to the BGK model results.

The Knudsen minimum for the even-velocity $2N+2$ hierarchy
is  observed for $N=10$, and the case $N=13$  i.e. $G$-$D1Q28\times D1Q3$ is presented in Fig. 6. The discrepancy between the model solution and the tabulated data  is approximately $10 \%$ for the Knudsen numbers in the part of the ballistic regime $1 \leq Kn \leq 3.33$. 
Similar  result can be  obtained  for an even-velocity high-order $D2Q4624$  off-lattice LB model with one relaxation time \cite{2011meng}. In the  present case the fully symmetric  $2D$ Gaussian LB model for $N=13$ in the $2N+2$ hierarchy has $28*28=784$ velocities ($G$-$D2Q784$). The  model $G$-$D2Q784$ is on-lattice and shorter than $D2Q4624$. Therefore, the Gaussian  method  shows faster convergence  to the benchmark results in ballistic regime than the  increase of the order  for the Gauss-Hermite quadratures in  conventional LB models.
\section{Conclusion}
The new discretization approach  for the kinetic BGK model is proposed.
This approach  somewhat intermediate for the classical LB method and DV approximation of the BGK model.
The  presented  hierarchy of the LB  has  several attractive  properties. The  streaming step is  linear and  the  method is conservative,  these properties are  inherited directly from the conventional LB models. Similarly to the DV methods for the BGK model the errors in the high order  moments can be  controlled  by the choice of the number of the summation steps.  Since  the shape of the equilibrium state after each summation step approaches closer to the Gauss distribution then the better reproduction of the kinetic boundary conditions is obtained. The  numerical experiments for the test problems (2D Couette and Poiseuille flows) support this fact.
Moreover, the summation monotonically enlarges the domain of positivity
for the local equilibrium state. This results in better stability, though the detailed investigation of stability properties should be  performed in future.
Finally, applying  the Central Limit Theorem several analytical properties are obtained: quadratic  convergence to the BGK model and  the closed form of the moment generating  function.

The presented construction confirms the idea that the precision of LB models (at least in the case of slow test flows) are mostly  influenced by the  structure of the lattice and the weights  but not the order of the lattice \cite{2016Succi_aip}. Moreover,  the results of the paper give an positive answer on the  question of the convergence  of the LB method to the BGK model  \cite{2016Succi_aip, 2007ansumali}.  Interestingly that this convergence  is achieved using  low-order lattices.

The  main drawbacks of the method  are the restriction to isothermal flows and  cubic growth  of the number  of lattice velocities for 3D problems.  This  features are  well-known for the conventional LB method and various techniques for overcoming of this difficulties are proposed. These questions are leaved for the future study.

\begin{figure}[h]
  \centerline{\includegraphics[width=0.45\textwidth, height=210pt]{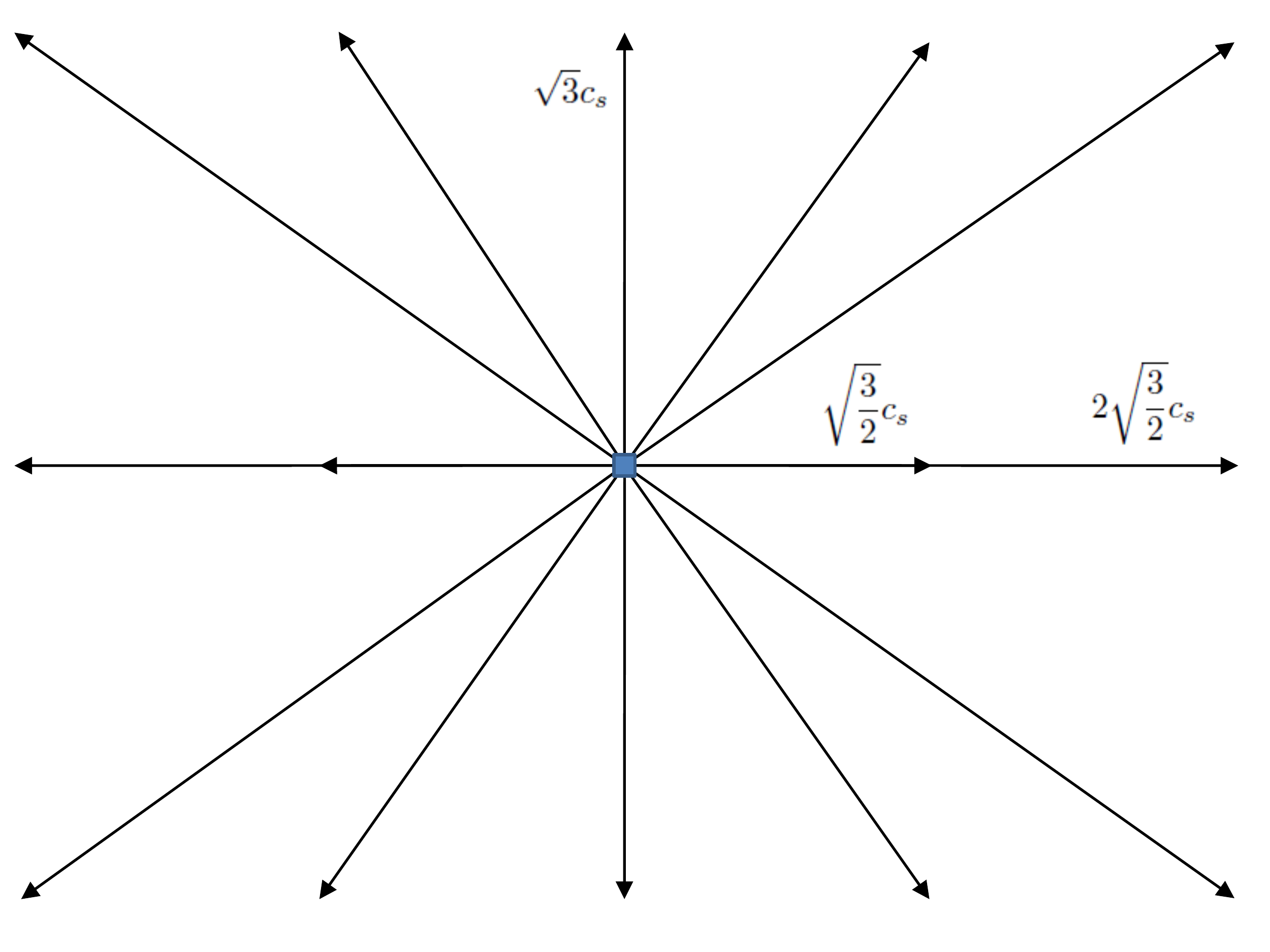}}
  \caption{ Lattice for $G$-$D1Q5 \times D1Q3$  model. 
  }
  \end{figure}
  
\begin{figure}[h]
  \centerline{\includegraphics[width=0.45\textwidth, height=210pt]{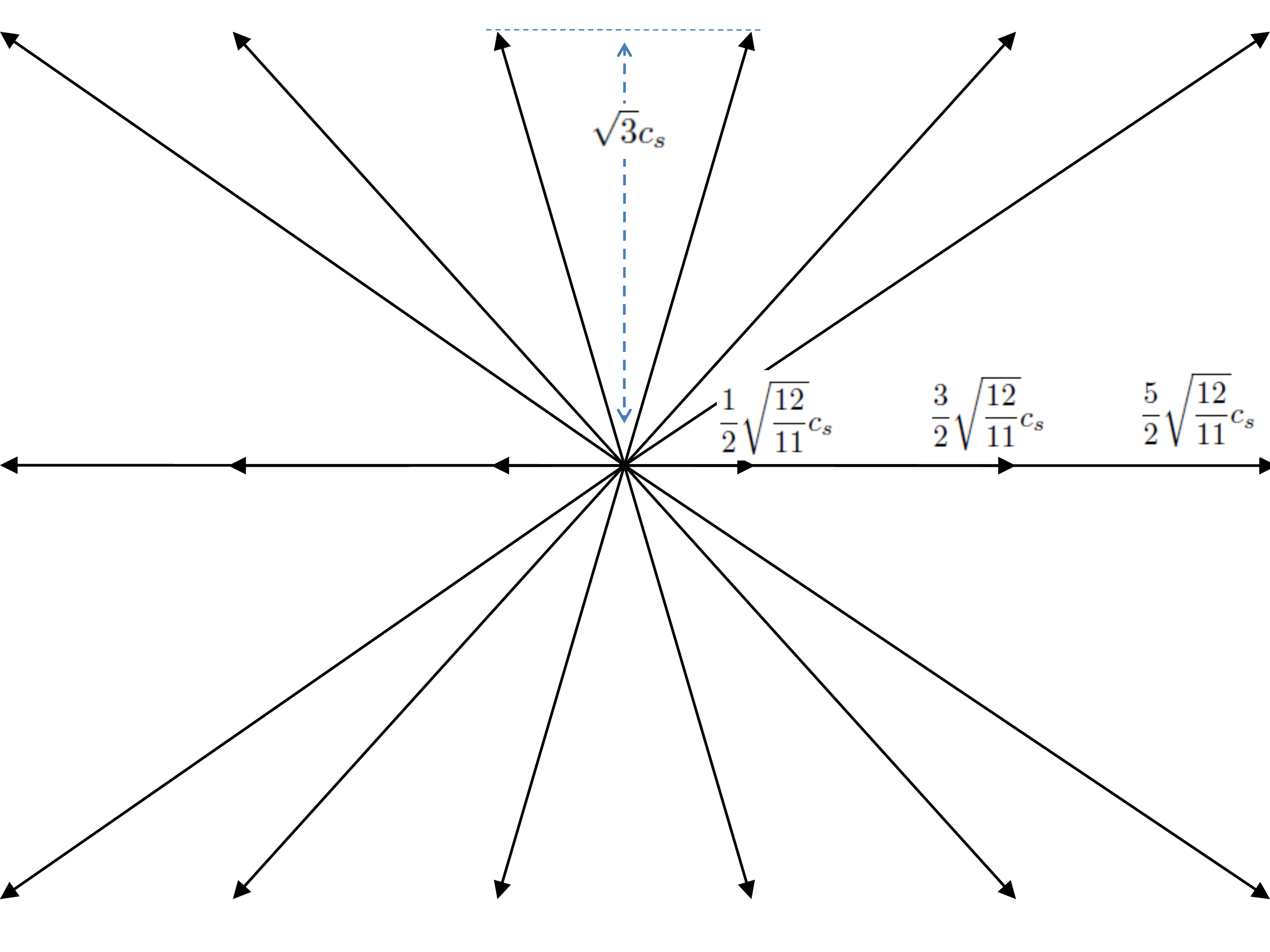}}
  \caption{ Lattice  $G$-$D1Q6 \times D1Q3$ model. 
  }
\end{figure}

\appendix

\section{$G$-$D1Q5=D1Q3+D1Q3$ example}

For a particular case of $N=2$ the $G$-$D1Q5$ (the formal sum  $D1Q3+D1Q3$) LB model uses $-2c,\, -c,\, 0,\, +c,\, +2c$ lattice and the following  local equilibrium state
is  obtained (here the particular case of unit temperature is considered)
$$
f_2^{eq}(2c)=\rho P_{2,+}^2, \quad
f_{1}^{eq}(c)=2\rho P_{2,0}P_{2,+}, 
$$
$$
f_{0}^{eq}(0)=\rho \bigl(P_{2,0}^2+2P_{2,+}P_{2,-}\bigr), 
$$
$$
f_{-1}^{eq}(-c)=2\rho P_{2,0}P_{2,-},  \quad
f_{-2}^{eq}(-2c)=\rho P_{2,-}^2, 
$$
where
$$
P_{2,\pm}=\frac{1}{4c^2}\left(1 \pm cu+\frac{u^2}{2}\right), 
$$
$$
P_{2,0}=1-P_{2,+}-P_{2,-} ,  \quad c=\sqrt{3/2}.
$$

When the bulk velocity $u$ is zero the  total equilibrium
state for $D1Q5$ has  the following form
$$
 f_2^{eq}(2c)=w_1^2, \quad
 f_{1}^{eq}(c)=2w_0w_1, 
$$
$$
f_{0}^{eq}(0)=w_0^2+2w_1^2, 
$$
$$
f_{-1}^{eq}(-c)=2w_0w_1, \quad
f_{-2}^{eq}(-2c)=w_1^2, 
$$
where $w_0=4/6, w_1=1/6$ are the  lattice weights for the $D1Q3$ model. This is  the simplest generalization of $D1Q3$ model. It should be  mentioned  that all the  models  in the hierarchy have  the same order of isotropy (as for $D1Q3$). This  is  unusual in comparison with the  conventional LB method where the  increase in number of discrete velocities leads to increase in number  of  exactly reproduced  moments.

\section{Third and Fourth moments}
Consider the  third moment $m_3$ for $2N+1$ hierarchy.
For the sake  of brevity in this Section it is assumed that $\rho=1$.
A simplest way to find the moments in an exact form is an application of  the  moment generating function (\ref{mgf}). One has for $2N+1$
$$
m_3^{(2N+1)}=\frac{d^3(P_{N,-}e^{-cs}+P_{N,0}+P_{N,+}e^{cs})^N }{ds^3}|_{s=0}=
$$
$$
=N(N-1)(N-2)(P_{N,+}-P_{N,-})^3+
$$
$$
+3N(N-1)(P_{N,+}-P_{N,-})(P_{N,+}+P_{N,-})+
N(P_{N,+}-P_{N,-})
$$
and remembering the  definitions of $P_{N,\pm}$ from  (\ref{maineq03}) the final result is obtained 
$$
m_3^{(2N+1)}=u^3+3u\theta_0-\frac{u^3}{N^2},
$$
where $\theta_0=c_s^2$ is the  gas temperature.

Another way to find this moment is based  on the fact  that the local equilibrium distribution  has the form of the  probability density for a sum of the independent and  identically distributed random variables, i.e. $Y=\sum_{j}^NX_j$. Then
$$
m_3^{(2N+1)}=\langle Y^3 \rangle=\left\langle\left(\sum_{j=1}^NX_j\right)^3\right\rangle
$$
and applying $\langle X_iX_jX_k \rangle=\langle X_i \rangle \langle X_j \rangle \langle X_k \rangle$ if $i \neq j \neq k$ one obtains the  expression
$$
m_3^{(2N+1)}=N(N-1)(N-2)\langle X \rangle^3+
3N(N-1)\langle X^2 \rangle\langle X \rangle+N\langle X^3 \rangle,
$$
where $X$ means any  of  $X_j$ . The latter expression leads to the same result.

The fourth moment can be  obtained from the moment generating function  or by computing $\langle Y^4 \rangle=\langle(\sum_{j}^NX_j)^4\rangle$, after some  lengthy algebra one obtains the following result  
$$
m_4^{(2N+1)}=u^4+6u^2\theta_0+3\theta_0^2+\left(\frac{4}{N^2}+\frac{3}{N^3}\right)u^4-\frac{3\theta_0 u^2}{N^2}.
$$

For $2N+2$ hierarchy the moment generating function was not obtained,  the third moment is then computed in a straightforward way
$$
m_3^{(2N+2)}=\frac{c^3}{2}\sum_{n=-N}^N Prob\left(\sum_j^NX_j=nc\right)
\times\left( \left(n-\frac{1}{2}\right)^3 +\left(n+\frac{1}{2}\right)^3 \right)=
$$
$$
=\sum_{n=-N}^N Prob\left(\sum_j^NX_j=nc\right)(nc)^3+\frac{3}{4}c^2u.
$$
and the sum in the expression above is the third moment for $2N+1$ hierarchy, then
$$
m_3^{(2N+2)}=u^3+3u\left(\theta_0+\frac{1}{4}c^2\right)-\frac{u^3}{N^2}
$$
finally remembering that the temperature for $2N+2$ hierarchy  is given by
 the relation (\ref{temp_2N2}) i.e. equals $\theta_0+\frac{c^2}{4}$ one can conclude that the error is again $u^2/N^3$
 in comparison with the local Maxwell third moment.

\section*{References}

\bibliography{main}

\end{document}